\documentclass[pra,10pt,twocolumn,superscriptaddress,showpacs]{revtex4-2}

\usepackage{nicefrac}
\usepackage{epsfig}
\usepackage{amsmath}
\usepackage{graphicx}
\usepackage{graphics}
\usepackage{float}
\usepackage{amssymb}
\usepackage[usenames,dvipsnames]{color}
\usepackage{natbib}
\usepackage{epstopdf}
\usepackage{verbatim}
\usepackage{wrapfig}
\usepackage{mathtools}
\usepackage{amsfonts}
\usepackage{dsfont}
\usepackage{bm}
\usepackage{physics}

\graphicspath{{Figure/}}

\makeatletter
\let\cat@comma@active\@empty
\makeatother

\begin{document}

\title{Improving the estimation of the environment parameters via a two-qubit scheme}

\author{Ali Raza Mirza}
\affiliation{School of Science \& Engineering, Lahore University of Management Sciences (LUMS), Opposite Sector U, D.H.A, Lahore 54792, Pakistan}

\author{Adam Zaman Chaudhry}
\email{adam.zaman@lums.edu.pk}
\affiliation{School of Science \& Engineering, Lahore University of Management Sciences (LUMS), Opposite Sector U, D.H.A, Lahore 54792, Pakistan}

\begin{abstract}

We demonstrate how using two qubits can drastically improve the estimation of environment parameters as compared to using only a single qubit. The two qubits are coupled to a common harmonic oscillatorenvironment, and the properties of the environment are imprinted upon the dynamics of the two qubits. The reduced density matrix of only one of these qubits contains a decoherence factor as well as an additional factor taking into account the indirect interaction induced between the qubits due to the interaction with their common environment. This additional factor can drastically improve the estimation of the environment parameters, as quantified by the quantum Fisher information. In particular, we investigate the estimation of the cutoff frequency, the coupling strength, and the temperature using our two-qubit scheme as compared to simply using a single qubit. For super-Ohmic environments in particular, one can improve the precision of the estimates by orders of magnitude. 

\end{abstract}


\maketitle

\section{Introduction}

Open quantum systems have attracted great interest over the past few decades as they have an important role in the development of modern quantum technologies \cite{nielsen2002quantum,haroche2014exploring}. In order to understand quantum dynamics properly, the effect of the environment must be taken into account as every quantum system inevitably interacts with its environment, leading to decoherence \cite{schlosshauer2007decoherence, breuer2002theory}. To understand the effect of decoherence, one must know parameters such as the system-environment coupling strength and the temperature. One useful method is to consider a quantum probe (a small and controllable quantum system) interacting with its environment \cite{brunelli2011qubit, brunelli2012qubit, neumann2013high, benedetti2014quantum,correa2015individual,elliott2016nondestructive,norris2016qubit,tamascelli2016characterization,streif2016measuring,benedetti2018quantum,cosco2017momentum,sone2017exact,salari2019quantum,razavian2019quantum,gebbia2020two,wu2020quantum,tamascelli2020quantum,gianani2020discrimination}. Once the dynamics of the probe are obtained, measurements on the probe allow us to estimate several properties associated with the environment. The precision of these estimates is encapsulated in the quantum Fisher information (QFI)  \cite{helstrom1976quantum,fujiwara2001quantum,monras2006optimal,paris2009quantum,monras2007optimal,genoni2011optical,spagnolo2012phase,pinel2013quantum,chaudhry2014utilizing,chaudhry2015detecting}. In order to minimize the error in the estimates, as dictated by the Cramer-Rao bound, one needs to maximize the QFI.

To date, many efforts have been put forward to estimate the environment parameters by using single-qubit and two-qubit quantum probes \cite{ benedetti2014quantum, benedetti2018quantum,benedetti2014characterization,wu2020quantum,tamascelli2020quantum}. In particular, the focus has been on estimating the parameters of a harmonic oscillator environment, taking the initial probe-environment state to be a simple product state. It has been argued that the use of a two-qubit probe is not really beneficial as compared to a single-qubit probe \cite{benedetti2018quantum}. By using a different scheme, in this paper, we argue on the contrary. Rather than performing measurements on two qubits to deduce the environment parameters, we show that the QFI can be drastically increased if we couple two qubits to the common environment, but perform measurements on only one of them. In other words, we trace out one of the qubits. The common environment induces an indirect interaction between the two qubits, and it is this indirect interaction (that depends on the environment parameters) that leads to the increase in the QFI. Moreover, if the environment is such that the probe-environment interaction is strong, then the assumption of weak probe-environment is questionable and the initial probe-environment correlations can become important \cite{hakim1985quantum,haake1985strong,grabert1988quantum,smith1990application,karrlein1997exact,
romero1997decoherence,lutz2003effect,banerjee2003general,van2004new,ban2009quantum,campisi2009fluctuation,
uchiyama2010role,dijkstra2010non,smirne2010initial,dajka2010distance,zhang2010different,tan2011non,
lee2012noncanonical,morozov2012decoherence,semin2012initial,chaudhry2013amplification,reina2014extracting,
chaudhry2013role,chaudhry2014effect,zhang2015role,chen2016effects,de2017dynamics,kitajima2017expansion,
buser2017initial,majeed2019effect,mirza2021master}. In this regard, it has been recently shown that these initial probe-environment correlations can also drastically increase the QFI \cite{ather2021improving}. In this spirit, we also take the initial correlations into account to show that, besides the indirect interaction, the initial correlations can further increase the QFI. Note that while we do trace out one qubit, we still consider our probe to be a two-qubit quantum probe as the presence of the second qubit is essential to drastically increase the QFI.    

We start by analyzing the dynamics of the two-qubit quantum probe interacting with a common harmonic oscillator environment. Assuming relaxation timescales to be much longer than the dephasing timescales, we ignore dissipation effects, that is, we consider pure dephasing. We allow our system and the environment interact each other until they achieve a joint thermal equilibrium state. A projective measurement is then performed on the probe to prepare the initial probe state. Thereafter, the two-qubit probe interacts with the environment. We then perform a partial trace over the environment as well as over the second qubit to obtain a $2\times2$ density matrix describing the dynamics of only qubit only. This density matrix contains the effect of decoherence, the initial correlations, as well as the indirect interaction between the two qubits. With this density matrix at hand, we work out the QFI, which is obviously a function of the interaction time between the probe and the environment. The idea then is to choose this interaction time such that the QFI is maximized. We conclusively show that the corresponding maximum QFI can be drastically greater than the QFI obtained with a single-qubit probe. 

This paper is organized as follows. In Sec.~\ref{model}, we present a detailed derivation of the reduced dynamics for our probe, both with and without initial correlations. In the next section, we compute the QFI and present results for the estimation of various environment parameters such as the cutoff frequency $\omega_c$, the coupling strength $G$, and the temperature $T$ respectively. In Sec.~\ref{optM}, we discuss the measurements that need to be performed in order to obtain the best estimates. Finally, in Sec.~\ref{ch5-sum}, we summarize our work. The appendices contain the detailed derivation of the dynamics. 

\section{the model}
\label{model}

We consider two qubits interacting with a common harmonic oscillator environment. We label the two qubits $1$ and $2$. The dynamics of our two-qubit system can be described by the Hamiltonian (we take $\hbar = 1$ throughout)
\begin{align}
    H
    =H_S+H_E+H_{SE}, \nonumber
\end{align}
where
\begin{align}
    H_S
    &= \frac{\omega_0}{2}\left( \sigma_z ^{\left(1\right)} + \sigma_z ^{\left(2\right)}\right),
\\
    H_E
    &= \sum_r \omega_r b_{r}^{\dagger} b_r,
\\
    H_{SE}
    &=\left( \sigma_z ^{\left(1\right)} + \sigma_z ^{\left(2\right)}\right)\sum_{r} \left(g_{r}^{*}b_r + g_r b_r^\dagger \right).
\end{align}
Here $\sigma_{x,y,z}$ are the Pauli spin operators, $\omega_0$ is the energy difference, $H_E$ is the harmonic oscillator environment Hamiltonian with the usual creation and annihilation operators (we have dropped the zero point energy for convenience), while $H_{SE}$ corresponds to the probe-environment interaction. To obtain the dynamics of the two qubits, it is convenient to first transform our Hamiltonian into an interaction picture via the unitary operator $U_0(t)=e^{-i \left(H_E+H_S\right) t}$. We then obtain
\begin{align*}
    H_{SE} \left(t\right)
    &=U_0 ^\dagger(t) H_{SE}  U_0(t),
\\
    &=\left( \sigma_z ^{\left(1\right)} + \sigma_z ^{\left(2\right)}\right) \sum_{r} \left(g_{r}^{*}b_r e^{-i\omega_r t} + g_r b_r^\dagger e^{i\omega_r t}\right).
\end{align*}
Using the Magnus expansion, this leads to the total unitary time-evolution operator being (see Appendix \ref{ch5-appA} for a detailed derivation)
\begin{align}
    U\left( t \right)  
    &=\text{exp}\left\{-i \left(\frac{\omega_0}{2}\left( \sigma_z ^{\left(1\right)} + \sigma_z ^{\left(2\right)}\right) + \sum_r \omega_r b_{r}^{\dagger} b_r \right)t\right\}\nonumber
\\    
    &\times \text{exp}\Bigg\{\frac{1}{2}\left( \sigma_z ^{\left(1\right)} + \sigma_z ^{\left(2\right)}\right) \sum_r\left[ \alpha_r \left(t\right)b_r^\dagger -  \alpha_r^* \left(t\right)b_r\right]\nonumber
\\    
    &-\frac{i}{2}\left(\mathds{1}+\sigma_z ^{\left(1\right)} \sigma_z ^{\left(2\right)}\right) \Delta \left( t \right)\Bigg\},
\end{align}    
with $\alpha_r\left( t \right)
    =\frac{2g_r\left(1-e^{i\omega_r t} \right)}{\omega_r}$, and  $\Delta \left( t \right) 
    =\sum_r   \frac{4\abs{g_r}^2}{\omega_r^2}\left[\sin(\omega_r t) - \omega_r t\right]$. 
    We first obtain the reduced density operator of the two-qubit probe via $\rho_S \left(t\right) 
        =\text{Tr}_E \left\{ U\left(t\right) \rho\left(0\right)U^\dagger\left(t\right) \right\}$.
It is useful to express this reduced density operator in the eigenbasis of $\sigma_z ^{\left(1\right)}$ and $\sigma_z ^{\left(2\right)},$ that is, $\ket{k,l}$, where $\sigma_z ^{\left(1\right)}\ket{k,l} =  k \ket{k,l}$ and $\sigma_z ^{\left(2\right)}\ket{k,l} =  l \ket{k,l}$. The two-qubit density matrix is then (the detailed calculation is presented in Appendix \ref{ch5-appB})
\begin{align}
    \left[\rho_S \left(t\right)\right]_{k',l';k,l} 
    &=e^{-i\frac{\omega_0}{2} \left(k'+l'-k-l\right)t} e^{-i\frac{\Delta\left(t\right)}{2} \left(k'l'-kl\right)}\nonumber
\\    
    &\times\text{Tr}_{\text{S,E}} \left\{  \rho \left(0\right) e^{-R_{kl,k'l'}\left(t\right)}P_{kl,k'l'} \right\}, \label{ch5-3} 
\end{align}
with $P_{kl,k'l'} \equiv \ket{k,l}\bra{k',l'}$, and 
\begin{align}
    R_{kl,k'l'}\left(t\right) 
    &=\sum_r \left[ \widetilde{\alpha}_r\left(t\right)b_r^\dagger -  \widetilde{\alpha}^*_r \left(t\right)b_r \right],
\\
    \widetilde{\alpha}_r\left(t\right) 
    &=\frac{1}{2}\left(k+l-k'-l'\right)\alpha_r\left(t\right).
\end{align}
\subsection{Factorized Initial State}
To make further progress, we now assume that the total state is a product state. In other words, denoting the initial state of the two qubits as $\rho_S(0)$ and the total state as $\rho(0)$, we have
\begin{align}
    \rho \left(0\right) 
    = \rho_S \left(0\right) \otimes \rho_E,
\end{align}
where $\rho_E 
= \frac{e^{-\beta H_E}}{Z_E} \  \text{with}  \  Z_E = \text{Tr}_E \left\{ e^{-\beta H_E} \right\}.$ From Eq.\eqref{ch5-3}, we then have 
\begin{align}
    \left[\rho_S \left(t\right)\right]_{k',l';k,l} 
    &=\left[\rho_S \left(0\right)\right]_{k',l';k,l}  \text{Tr}_{E} \left\{ \rho_Ee^{-R_{kl,k'l'}\left(t\right)}\right\} \nonumber
\\    
    &\times e^{-\frac{i \omega_0}{2} \left(k'+l'-k-l\right)t} e^{-\frac{i\Delta\left(t\right)}{2} \left(k'l'-kl\right)}.\label{ch5-4}
\end{align}
We now simplify $\text{Tr}_{E} \left\{ \rho_Ee^{-R_{kl,k'l'}\left(t\right)}\right\}$. Since the modes of the harmonic oscillator are independent of each other, we obtain
\begin{align}
    \expval{e^{-R_{kl,k'l'}\left(t\right)}}
    &= \prod_r  \text{exp} \left\{-\frac{1}{2}\abs{\widetilde{\alpha}_r\left(t\right)}^2 \expval{2n_r + 1 }\right\},
\end{align}
where we have defined $n_r = \expval{b_r^\dagger b_r}$. The environment being in thermal equilibrium, $n_r$ is simply the Bose-Einstein distribution, that is, $n_r =  \frac{1}{e^{{\beta\omega_r }} -1}=\frac{1}{2}\left\{\coth\left(\frac{\beta \omega_r}{2}\right)-1\right\}$. Therefore
\begin{align}
    \text{Tr}_{E} \left\{ \rho_Ee^{-R_{kl,k'l'}\left(t\right)}\right\}
    &= \text{exp}\left\{ -\frac{1}{4}\left(k+l-k'-l'\right)^2\Gamma\left(t\right)\right\}\nonumber,
\end{align}
with
\begin{align}
    \Gamma\left(t\right)
    =&\sum_r  \frac{ 4\abs{g_r}^2}{\omega_r^2} \left[1-\cos\left(\omega_r t \right)\right]\coth{\left( \frac{\beta \omega_r}{2}\right)}.
\end{align}
To sum up, the final state of the two-qubit probe can be written
\begin{align}
    &\left[\rho_S \left(t\right)\right]_{k',l';k,l} 
    =\left[\rho_S \left(0\right)\right]_{k',l';k,l}\nonumber
\\    
    &\times e^{-i\frac{\omega_0}{2} \left(k'+l'-k-l\right)t} e^{-i\frac{\Delta\left(t\right)}{2} \left(k'l'-kl\right)}e^{ -\frac{1}{4}\left(k+l-k'-l'\right)^2\Gamma\left(t\right)}\nonumber.
\end{align}
Note that $\Gamma\left(t\right)$ describes decoherence, while $\Delta(t)$ describes the indirect interaction between the qubits due to the interaction with the common environment. We take the initial state to be ‘pointing up’ along the $x$-axis, that is, $\rho_S\left(0\right)=\ket{+,+}\bra{+,+},$ where $  \sigma_x\ket{+}=\ket{+}$. 
Also, the effect of the environment on the system is encapsulated by the spectral density of environment $J(\omega)$. This function effectively converts a sum over the environment modes to an integral via $\sum_r 4 \abs{g_r}^2 f\left(\omega_r\right) \rightarrow \int_0^\infty d\omega \, J\left(\omega\right) f\left(\omega\right)$. We consider the spectral density is usually assumed to be of the form 
$J\left(\omega\right)
= G\frac{\omega^s}{\omega_c^{s-1}}e^{\frac{\omega}{\omega_c}}$ \cite{schlosshauer2007decoherence}. Here $G$ is the coupling strength, $\omega_c$ is the cutoff frequency, and $s$ is the Ohmicity parameter with $s<1$, $s=1$ and $s>1$ representing sub-Ohmic, Ohmic, and super-Ohmic spectral densities respectively. By taking a partial trace over the second qubit, the state of the first qubit alone is obtained as 
\begin{align}
\rho_{\text{S1}}^\text{un}\left(t\right)
&=
\begin{pmatrix}
    1/2 
    & \frac{e^{-i\omega_0 t - \Gamma_{\text{un}}\left(t\right)}\cos\left[\Delta\left(t\right)\right]}{2}\\
    \frac{e^{i\omega_0 t - \Gamma_{\text{un}}\left(t\right)}\cos\left[\Delta\left(t\right)\right]}{2}
    & 1/2
\end{pmatrix}\nonumber,
\end{align}
with
\begin{align*}
    \Gamma_\text{un}\left(t\right)
    &=\int_0^\infty J(\omega)   \left[1-\cos\left(\omega t \right)\right]\coth{\left( \frac{\beta \omega}{2}\right)d\omega},
\\
    \Delta\left(t\right)
    &=\int_0^\infty \frac{J(\omega)}{\omega^2} \left[\sin\left(\omega t \right)-\omega t\right] d\omega.
\end{align*}
Note that we have added the subscript `un' to emphasize that this is the decoherence factor when we have an uncorrelated initial state. It is useful to split $\Gamma_{\text{un}}(t)$ into temperature-dependent and temperature-independent parts, that is, $\Gamma_{\text{un}}(t) = \Gamma_{\text{vac}}(t) + \Gamma_{\text{th}}(t)$ \cite{morozov2012decoherence}. At zero temperature, $\Gamma_{\text{th}}(t) = 0$. On the other hand
\begin{align*}
    &\Gamma_{\text{vac}}(t)
\\    
    &=\begin{cases}
    \frac{G}{2}\ln\left(1+ \omega^2_c t^2\right)  &  s=1,
\\    
    G \Bar{\Gamma}[s-1] -  \frac{1}{2}\left(\frac{G \Bar{\Gamma}[s-1]}{\left(1 + i\omega_c t\right)^{s-1}} + \frac{G \Bar{\Gamma}[s-1]}{\left(1+i\omega_c t\right)^{s-1}}\right)  &  s\neq 1,
    \end{cases}
\end{align*}
where $\bar{\Gamma}$ is the usual gamma function defined as $\bar{\Gamma}[z]=\int_{0}^{\infty} t^{z-1} e^{-t} dt$.
\subsection{Correlated Initial State}
We now consider preparing the initial state of the two qubits in the initial state $\ket{\psi} = \ket{+,+}$ via a projective measurement. The total system environment initial state is then written as

\begin{align}
    \rho \left(0\right) 
    =\ket{\psi}\bra{\psi} \otimes \frac{\bra{\psi} e^{-\beta H} \ket{\psi}}{Z}, \label{ch5-7}
\end{align}
where 
    $Z =  \text{Tr}_{\text{S,E}}\left\{ e^{-\beta H} \right\}$ is the total partition function. Inserting the completeness relation $\sum_{p,q} \ket{p,q}\bra{p,q} $ such that $\sigma^{(1)}_z \ket{p,q} = p \ket{p,q}$ and $\sigma^{(2)}_z \ket{p,q} = q \ket{p,q}$, and using the displaced harmonic oscillator modes 
    $B_{r,p,q}
    = b_r + \frac{\left( p+q\right)g_r}{\omega_r}$, it is straightforward to show that
\begin{align}
    Z
    &=\sum_{p,q}e^{-\frac{\beta\omega_0}{2}\left( p+q\right)}e^{\beta \sum_r(p+q)^2\frac{\abs{g_r}^2}{\omega_r}
    }Z_E.
\end{align}
After algebraic manipulations, we arrive at the final expression of $\rho_S \left(t\right)$, namely 
\begin{align}
    &\left[\rho_S \left(t\right)\right]_{k',l';k,l} 
    =\left[\rho_S \left(0\right)\right]_{k',l';k,l}X\left(t\right)\nonumber
\\    
    &\times e^{-i\frac{\omega_0}{2} \left(k'+l'-k-l\right)t} e^{-i\frac{\Delta\left(t\right)}{2} \left(k'l'-kl\right)}e^{ -\frac{1}{4}\left(k+l-k'-l'\right)^2\Gamma\left(t\right)}  \nonumber,
\end{align}
where
\begin{align}
    &\left[\rho_S \left(0\right)\right]_{k',l';k,l} 
    =\bra{\psi}  \ket{k,l}\bra{k',l'}\ket{\psi},\nonumber
\\
    X\left(t\right)
    &=\frac{\sum_{p,q}e^{-\frac{\beta\omega_0}{2}\left( p+q\right)}\abs{\bra{p,q}\ket{\psi}}^2e^{\beta(p+q)^2 \frac{\mathcal{C}}{4}}e^{-i(p+q)\widetilde{\Phi}\left(t\right)} }{\sum_{p,q}e^{-\frac{\beta\omega_0}{2}\left( p+q\right)}\abs{\bra{p,q}\ket{\psi}}^2e^{\beta(p+q)^2 \frac{\mathcal{C}}{4}}},\nonumber
\end{align}
$\mathcal{C}
    =\sum_r \frac{4\abs{g_r}^2}{\omega_r}$ and  $\widetilde{\Phi}_{k',l';k,l}\left(t\right) 
    =\frac{1}{2}\left(k+l-k'-l'\right)\phi\left(t\right)$
with
\begin{align}
    \phi\left(t\right)
    =\int_0^\infty G\frac{\omega^s}{\omega_c^{s-1}}e^{-\frac{\omega}{\omega_c}} \frac{\sin\left(\omega t \right)}{\omega^2}d\omega.
\end{align} 
From this state, we get the state describing the dynamics of the first spin system by taking a partial trace over the second spin system, as we did in the uncorrelated case. We write the final result as 
\begin{align}
    \rho_{\text{S1}}^\text{corr} \left(t\right)
    &=\begin{pmatrix}
    1/2 
    & \frac{e^{-i\xi\left(t\right)-\Gamma\left(t\right)}\cos\left[\Delta\left(t\right)\right]}{2} 
\\
    \frac{e^{i\xi\left(t\right)-\Gamma\left(t\right)}\cos\left[\Delta\left(t\right)\right]}{2}
    & 1/2
\end{pmatrix}, \label{ch5-1}
\end{align}
where $\xi\left(t\right)=\omega_0 t + \chi\left(t\right)$. Again, $\Gamma\left(t\right)$ incorporates the decoherence effect of the environment, while $\Delta(t)$ captures the indirect interaction. Moreover, the effect of the initial correlations emerges as an effective level shift $\chi\left(t\right)$ as well as a modification of the decoherence factor. In particular, the decoherence factor is now
\begin{align}
    \Gamma\left(t\right)
    &=\Gamma_\text{un}\left(t\right)+\Gamma_\text{corr}\left(t\right),\nonumber
\\
    \Gamma_\text{corr}\left(t\right)
    &=\text{ln}\left[\frac{1+e^{\beta \mathcal{C}}\cosh\left(\beta \omega_0\right)}{\sqrt{a^2 \left(t\right) + b^2 \left(t\right) }}\right],\label{ch5-12}
    \end{align}
    while the effective level shift is
    \begin{align}
    \chi\left(t\right)
    &=\text{tan}^{-1}
    \left[\frac{b \left(t\right)}{a \left(t\right)}\right].\nonumber
\end{align}
Here we have defined the time-dependent coefficients $a \left(t\right) = 1+e^{\beta \mathcal{C}}\cosh\left(\beta \omega_0\right)\cos[2\phi\left(t\right)]$ and $b \left(t\right) = e^{\beta \mathcal{C}}\sinh\left(\beta \omega_0\right)\sin[2\phi\left(t\right)]$, with
\begin{align}
    \phi(t)=
    \begin{cases}
    G\, \text{tan}^{-1}\left(\omega_c t\right) & \quad s=1,
\\
    \frac{G}{2i} \left(\frac{1}{\left(1 - i\omega_c t\right)^{s-1}} - \frac{1}{\left(1+i\omega_c t\right)^{s-1}}\right)\Bar{\Gamma}[s-1] & \quad s\neq 1.
    \end{cases}\nonumber
\end{align}

\section{Quantum Fisher Information} \label{QFI}
With the dynamics at hand, both with and without the initial correlations, we now move to calculate the quantum Fisher information (QFI). The QFI quantifies the precision with which a general environment parameter $\textit{x}$ can be estimated \cite{benedetti2018quantum}. It can be shown that the QFI is related to the Cramer-Rao bound - the greater the QFI, the greater our precision of the estimate. The general expression for the QFI is given by
\begin{align}
    \mathds{F}_{Q}\left(x\right)
    &=\sum_{n=1}^2 \frac{(\partial_x\rho_n)^2}{\rho_n}+2\sum_{n\neq m }\frac{(\rho_n-\rho_m)^2}{\rho_n+\rho_m}\abs{\bra{\varepsilon_m}\ket{\partial_x\varepsilon_n}}^2,\label{ch5-2}
\end{align}
where $\ket{\varepsilon_n}$ is the $n^{\text{th}}$ eigenstate of our reduced single qubit state and $\rho_n$ is the corresponding eigenvalue. For the evaluated single qubit $2 \times 2$ matrix, it is straightforward to calculate the eigenvalues and eigenstates. We find that $\rho_1=\frac{1}{2}[1-\mathcal{F}\left(t\right)] $ and $\rho_2=\frac{1}{2}[1+\mathcal{F}\left(t\right)] $ with $\mathcal{F}\left(t\right)= \cos\left[\Delta(t)\right]e^{-\Gamma\left(t\right)}$. The corresponding eigenstates are
\begin{align*}
    \ket{\epsilon_1\left(t\right)}
    &=\frac{1}{\sqrt{2}}\left\{\ket{0} + e^{i\xi \left(t\right)}\ket{1}\right\},
\\
    \ket{\epsilon_2\left(t\right)}
    &=\frac{1}{\sqrt{2}}\left\{\ket{0} - e^{i\xi \left(t\right)}\ket{1}\right\},
\end{align*}
where $\ket{0}$ and $\ket{1}$ being the eigenstates of $\sigma_z$ and following the eigenvalue equation $\sigma_z\ket{n}=(-1)^n\ket{n}$. Now
\begin{align}
    \left(\partial_x \rho_1 \right)^2
    =\left(\partial_x \rho_2 \right)^2
    &=\frac{1}{4}e^{-2\Gamma} \left(\sin \Delta \partial_x \Delta + \cos \Delta \partial_x \Gamma \right)^{2}.\nonumber 
\end{align}
Calculating also the derivatives of the eigenstates, and substituting in Eq.~\eqref{ch5-2}, the QFI comes out to be
\begin{align}
    \mathds{F}_{Q}\left(x\right)
    &=\frac{\left( \sin{\Delta} \partial_x \Delta  + \cos{\Delta} \partial_x\Gamma \right)^2}{e^{2\Gamma}-\cos^2{\Delta}} + \frac{\cos^2{\Delta}\left(\partial_x \chi\right)^2}{e^{2\Gamma}}.\label{ch5-5}
\end{align}
This expression reduces to the expression presented in Ref. \cite{ather2021improving} for a single qubit case by setting $\Delta = 0$. Eq. \eqref{ch5-5} gives the QFI for the case where we take the initial correlations into account. If we start with the simple product, then we can obtain the QFI by setting $\chi = 0$ and replacing $\Gamma$ by $\Gamma_{\text{un}}$.

Before moving on to concrete examples of estimating the environment parameters, let us note that the addition of the indirect qubit-qubit interaction appears highly promising in increasing the QFI, in particular due to the presence of the $\sin \Delta \partial_x \Delta$ term in the QFI.  

\subsection{Estimation of the cutoff frequency of the environment}
As the first example of the application of our expression of the QFI (see Eq.~\ref{ch5-5}), we now look in detail at the estimation of the cutoff frequency $\omega_c$ of the environment. We first note that 
\begin{figure}[t]
 		\includegraphics[scale = 0.8]{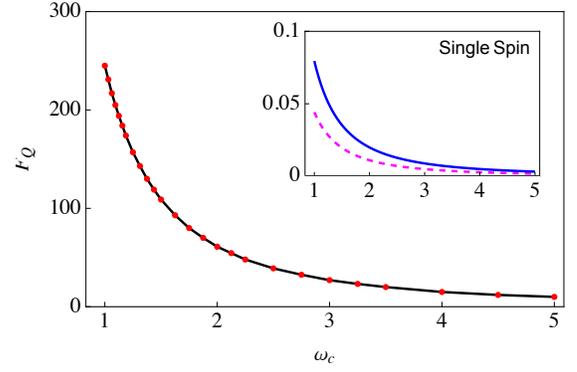}
 		\centering
		\caption{(Color online) The main figure shows the behavior of the optimized QFI for the estimation of the cutoff frequency as a function of the cutoff frequency. The black, solid curve is obtained by including the effects of the initial correlations, while the dotted, red curve ignores these effects. We have taken $\omega_0 = 1$ and the rest of the parameters are  $G= 0.01$, $s=0.5$, and the temperature $T = 0$. The inset shows the optimized QFI if we simply use a single qubit or spin (without any second qubit), both with (solid, blue curve) and without correlations (dashed, magenta curve). The parameters used are the same as the main figure.}
		\label{weakcoupling}
\end{figure}
\begin{figure}[t]
 		\includegraphics[scale = 0.8]{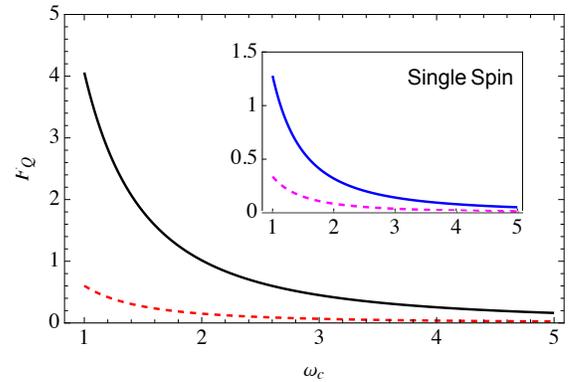}
 		\centering
		\caption{(Color online) Same as Fig. \ref{weakcoupling}, except that $G=1$.}
		\label{StrongCoupling}
\end{figure}
\begin{figure}[t]
 		\includegraphics[scale = 0.8]{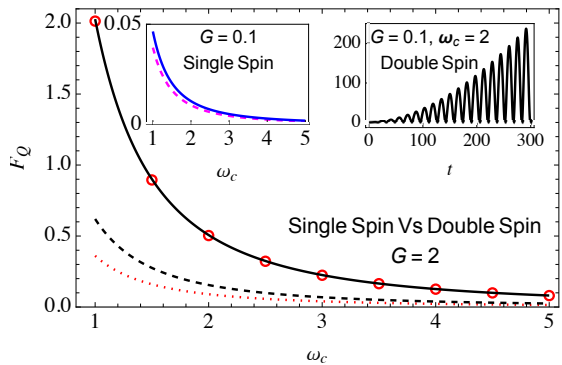}
 		\centering
		\caption{(Color online) Comparison of the optimized QFI for estimating $\omega_c$ for a single qubit probe versus our two-qubit scheme for an Ohmic environment $(s=1)$. In the main plot, the solid black (with initial correlations) and dashed black (without correlations) show the optimized QFI for the two-qubit scheme while the red circles (with correlations) and dotted red (without correlations) show the optimized QFI for the single-qubit case. In the top-left inset, the optimized QFI is plotted with (solid blue) and without (magenta dashed) correlations with $G=0.1$ for the single-qubit case while in the top-right inset, the QFI is plotted with (solid black) and without (red dashed) correlations at $G=0.1$ for the two-qubit scheme. Other parameters are the same as Fig. \ref{weakcoupling}.}
		\label{ohmic}
\end{figure}
\begin{figure}[t]
 		\includegraphics[scale = 1]{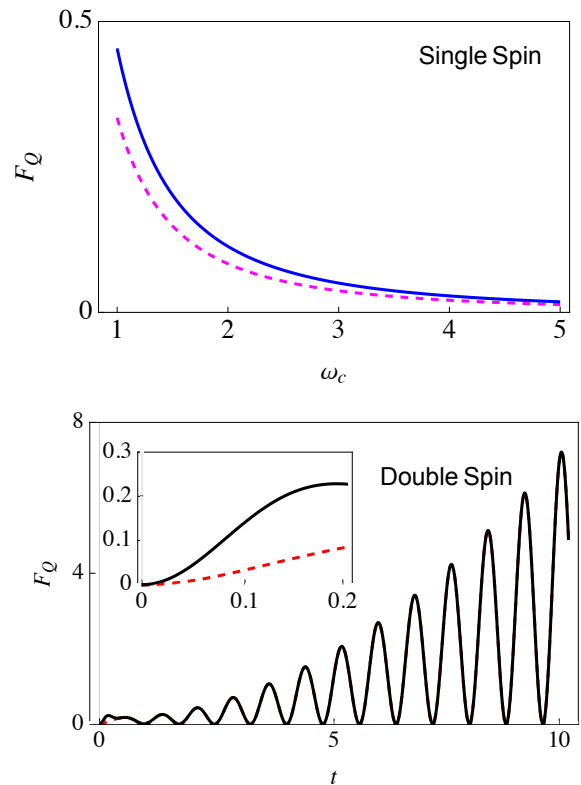}
 		\centering
		\caption{(Color online) We now consider a super-Ohmic environment with $s = 2$. The top plot shows the optimized QFI for estimating $\omega_c$ with (solid blue) and without (dashed magenta) correlations using a single qubit as the probe. The bottom plot shows the QFI with (solid black) and without (dashed red) correlations with the two-qubit scheme. The inset simply zooms in on the QFI for small values of time to illustrate the QFI with and without correlations. Other parameters used are the same as Fig.~\ref{weakcoupling} except that now we have $G=2$.}
		\label{Super1}
\end{figure}
\begin{align*}
\frac{\partial\Gamma}{\partial \omega_c}
&=\begin{cases}
    \frac{G \omega_c t^2}{1 + \omega^{2}_c t^2} & \quad s = 1,
\\    
    iG \Bar{\Gamma}[s]t\left(\frac{1}{2\left(1+i\omega_c t\right)^s} - \frac{1}{2\left(1-i\omega_c t\right)^s}\right) & \quad s \neq 1,
\end{cases}
\\
    \frac{\partial\Delta}{\partial \omega_c}
    &=\begin{cases}
    \frac{G \omega_c^2 t^3}{1 + \omega^{2}_c t^2} & \hspace{0.18cm} s = 1,
\\    
    G \Bar{\Gamma}[s]t\left(\frac{1/2}{\left(1+i\omega_c t\right)^s} + \frac{1/2}{\left(1-i\omega_c t\right)^s} - 1 \right) & \hspace{0.18cm} s \neq 1,
\end{cases}    
\\
\frac{\partial\chi}{\partial \omega_c}
&=\begin{cases}
    \frac{2G t}{1 + \omega^{2}_c t^2} & \hspace{0.5cm} s = 1,
\\    
    -G \Bar{\Gamma}[s]t\left(\frac{1}{\left(1+i\omega_c t\right)^s} + \frac{1}{\left(1-i\omega_c t\right)^s}\right) & \hspace{0.5cm} s \neq 1.
\end{cases}
\end{align*}
Using these in Eq.~\eqref{ch5-5}, we obtain the quantum Fisher information for the estimation of the cutoff frequency as a function of time. We then optimize this QFI over the interaction time to find the maximum possible QFI. For example, one could plot the QFI as a function of time for different values of $\omega_c$, and thereby note the maximum value of QFI for each value of $\omega_c$. We can then investigate the behavior of this optimal QFI as a function of the cutoff frequency. This behavior is illustrated in Fig.~\ref{weakcoupling}. The main figure shows the typical behavior of the QFI for estimating the cutoff frequency for a sub-Ohmic environment using our two-qubit scheme, both with and without including the effect of the initial correlations. It is clear that in this weak coupling strength regime, the effect of the initial correlations is insignificant, as expected since the black, solid curve overlaps with the red dotted curve. The inset shows the optimized QFI if we simply use a single qubit interacting with the environment with the same set of parameters. What is most notable in this figure is the drastic increase of the QFI with our two-qubit scheme as compared to using a single qubit - it is a three orders of magnitude increase, which demonstrates in a remarkable manner the advantage of using our two-qubit scheme. The increase is simply because of the indirect qubit-qubit interaction (the $\Delta$ term). Interestingly, if we increase the coupling strength $G$, our two-qubit scheme improves the QFI, although the increase is not as drastic as the in the case of weak coupling (see Fig.~\ref{StrongCoupling}) - the increased decoherence leads to the smaller values of the QFI. We also investigated an Ohmic environment in Fig.~\ref{ohmic}. For strong coupling, we notice the overlap of red circles (using the simple single qubit probe with correlations included) and the solid black curve (using our two-qubit scheme with the effect of the correlations included), thereby indicating that the two schemes perform similarly for strong coupling with an Ohmic environment. However, the situation drastically changes for weaker coupling. As one can see from the inset, the QFI with our two-qubit scheme keeps on increasing as the qubits interact with their environment - the decoherence is now smaller, and the indirect interaction leads to a buildup of the information gained about the environment. On the other hand, the QFI obtained using a single qubit probe is bounded. Similar behavior is seen with a super-Ohmic environment (see Fig.~\ref{Super1}), where again the indirect inter-qubit interaction (the $\Delta$ term) plays a vital role in increasing the QFI. In fact, now the buildup of QFI with the two-qubit scheme persists even in the strong coupling regime. 
\subsection{Estimation of system-environment coupling strength}
We now consider estimating the coupling strength $G$. We again use the expression given in Eq.~\eqref{ch5-5} and optimize it over the interaction time to get optimized QFI. We need now the derivatives
\begin{align*}
\frac{\partial\Gamma}{\partial G}
&=\begin{cases}
    \frac{1}{2}\ln \left(1 + \omega^{2}_c t^2\right) &  s = 1,
\\    
    \Bar{\Gamma}[s-1] - \left(\frac{\Bar{\Gamma}[s-1]/2}{\left(1 + i\omega_c t\right)^{s-1}} + \frac{\Bar{\Gamma}[s-1]/2}{\left(1 + i\omega_c t\right)^{s-1}} \right) &  s \neq 1,
\end{cases}
\\
\frac{\partial\Delta}{\partial G}
    &=\begin{cases}
    \text{tan}^{-1}\left(\omega_c t\right) - \omega_c t &  \hspace{0.1cm} s=1,
\\
    \Bar{\Gamma}[s] \omega_c t - \left(\frac{i\Bar{\Gamma}[s-1]/2}{\left(1-i\omega_c t\right)^{s-1}} -  \frac{i\Bar{\Gamma}[s-1]/2}{\left(1 + i\omega_c t\right)^{s-1}} \right) & \hspace{0.1cm} s\neq 1,
    \end{cases}
\\
\frac{\partial\chi}{\partial G}
    &=\begin{cases}
    2\text{tan}^{-1}\left(\omega_c t\right) & \hspace{0.2cm} s=1,
\\
    i\Bar{\Gamma}[s-1] \left( \frac{1}{\left(1 - i\omega_c t\right)^{s-1}} - \frac{1}{\left(1+i\omega_c t\right)^{s-1}}\right) & \hspace{0.2cm} s\neq 1.
    \end{cases}
\end{align*}
\begin{figure}[t]
 		\includegraphics[scale = 0.8]{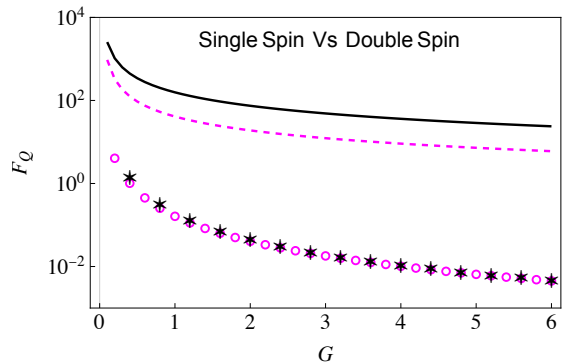}
 		\centering
		\caption{(Color online) Behavior of optimized QFI versus coupling strength $G$ obtained using a single qubit probe [magenta with (dashed) and without (circles) initial correlations], and our two-qubit scheme [black with (solid) and without (asterisks) correlations]. Here we have considered a sub-Ohmic ($s=0.1$) environment. Also, $\omega_{c}=5$, with the rest of the parameters the same as in Fig. \ref{weakcoupling}.}
		\label{HvsA}
\end{figure}
\begin{figure}[t]
 		\includegraphics[scale = 0.8]{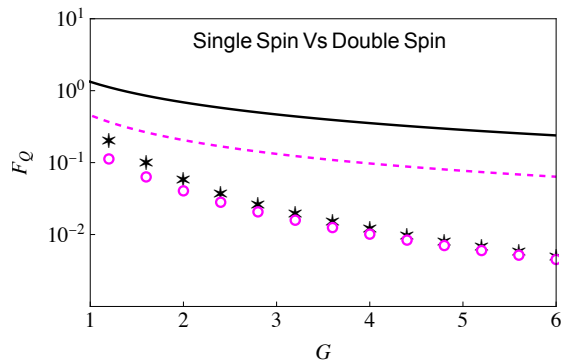}
 		\centering
		\caption{(Color online) Same as Fig.~\ref{HvsA}, except that now we considering an Ohmic environment ($s=1$).}
		\label{GOhmic}
\end{figure}
\begin{figure}[t]
 		\includegraphics[scale = 1]{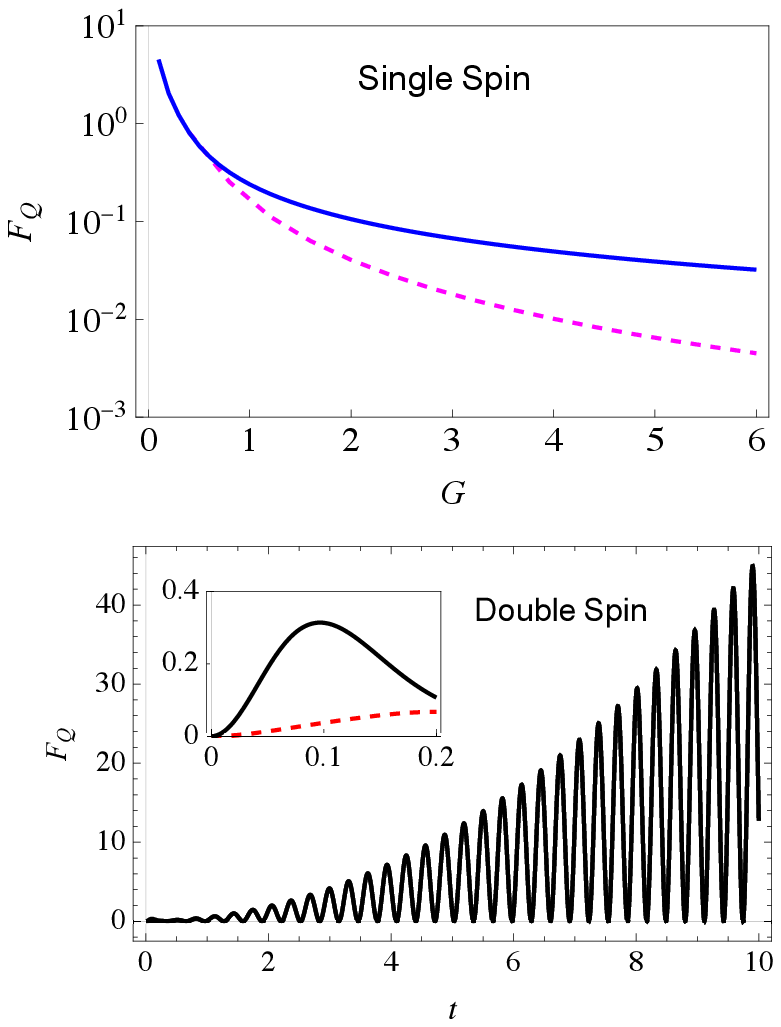}
 		\centering
		\caption{(Color online) Same as Fig.~\ref{Super1}, but here we are estimating the coupling strength. Also, we are using $\omega_c = 5$.}
		\label{GSuper}
\end{figure}
We first compare the optimized QFI for estimating the coupling strength $G$ obtained using our two-qubit scheme with the QFI obtained using a single-qubit probe for a sub-Ohmic environment. Results are illustrated in Fig.~\ref{HvsA}, where we have shown the behavior of the optimized QFI versus the coupling strength $G$ using a single qubit probe both with and without incorporating the effect of the initial correlations - these are shown with the dashed, magenta curve and the circular markers respectively. We have also shown the QFI with our two-qubit scheme, both with  (solid, black curve) and without (the asterisk markers) including the initial correlations. At least three points should be noted here. First, if we ignore the initial correlations, then there is little difference between the two schemes. Second, the role of the initial correlation is, in general, very important. Third, with both indirect interactions and the initial correlations accounted for, there is a drastic increase in the QFI. Following the same color scheme and parameters used in Fig. \ref{HvsA}, we demonstrate the optimized QFI in an Ohmic environment $s=1$ as well (see Fig. \ref{GOhmic}). In this environment, while the QFI is lower as compared to the sub-Ohmic environment, the benefit of using our two-qubit scheme is still evident.

The advantage of our two-qubit scheme becomes even more evident, as before, with super-Ohmic environments (see Fig.~\ref{Super1}). Once again, the QFI generally keeps on increasing as we increase the interaction time (see the bottom figure in Fig.~\ref{GSuper}) for our two-qubit scheme. If we compare this with the results obtained using a single qubit probe with (solid blue curve) or without (magenta dashed curve) correlations (see the top plot), we see that the QFI for the single qubit probe is far smaller.

\subsection{Estimation of Temperature}

\begin{figure}[t]
 		\includegraphics[scale = 0.8]{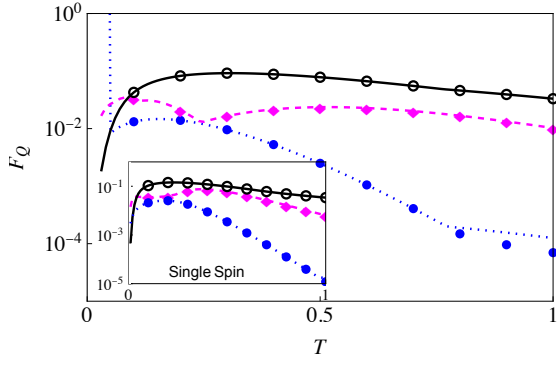}
 		\centering
		\caption{(Color online)  The QFI for the estimation of temperature. The black curves [with (dashed) and without  (circles) correlations], magenta curves [with (solid) and without  (squares) correlations], and blue curves [with (dotted) and without  (solid circles) correlations] denote the optimized QFI with a super-Ohmic $(s=2)$, Ohmic $(s=1)$ and sub-Ohmic $(s=0.5)$ environment respectively. Here we have $\omega_c=5$ and $G=1$. The inset follows the same parameters and color scheme but for the single qubit probe.}
		\label{TempAll}
\end{figure}

We now consider the estimation of temperature using a single qubit probe, as well as using our two-qubit scheme, for sub-Ohmic, Ohmic, and super-Ohmic environments. Since temperature is not zero here, therefore $\Gamma_{\text{corr}}\left(t\right)$ and $\Gamma_{\text{th}}(t)$ are no longer zero. $\Gamma_{\text{corr}}\left(t\right)$ can be found analytically -  its expression is given in Eq.~\eqref{ch5-12} - while $\Gamma_{\text{th}}(t)$ and its temperature derivative are found numerically. We illustrate our results in Fig.~\ref{TempAll}. The key point to note here is that the higher temperatures mean that the decoherence factor is greatly enhanced. This enhancement effectively washes out the advantage of using our two-qubit scheme, so that the quantum Fisher information with a single-qubit probe and our two-qubit scheme are quantitatively similar.  

\begin{figure}[t]
 		\includegraphics[scale = 0.8]{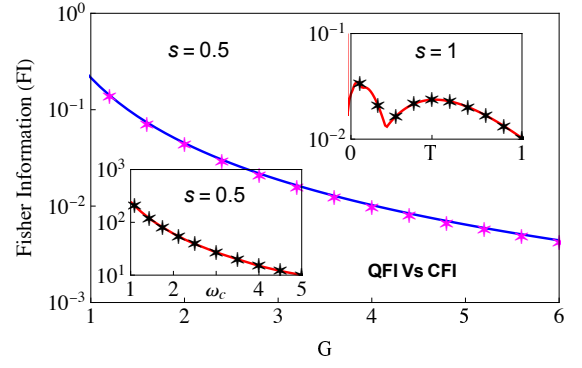}
 		\centering
		\caption{(Color online) Plot of QFI (solid curves) versus CFI (asterisk markers). The main plot shows the estimation of the coupling strength $G$ with $\omega_c=5$. In the insets, we have plotted the optimized Fisher information (quantum and classical). On the top right, we are estimating the temperature $T$ with coupling strength $G=1$. At the bottom left, we estimate the cutoff frequency with coupling strength $G=0.01$.}\label{Q3D}
\end{figure}

\section{OPTIMAL MEASUREMENT}
\label{optM}

Until now, we have found that by using our two-qubit scheme, the QFI is substantially increased. The question remains regarding which measurements need to be performed in order to obtain this maximum QFI. This can be answered by calculating the classical Fisher information (CFI) for a particular measurement scheme; if the CFI comes out to be equal to the QFI, then we have found the optimal measurement to be performed. We guess that the optimal measurements are projective measurements described by the projection operators $\mathbf{P}_1 = \ket{\Psi_1}\bra{\Psi_1}$ and $\mathbf{P}_2 = \ket{\Psi_2}\bra{\Psi_2}$, with
\begin{align}
    \ket{\Psi_1} 
    &=\frac{1}{\sqrt{2}}\left\{
    \ket{\uparrow}_{z}
+ e^{i\varphi}\ket{\downarrow}_{z}\right\},
\\
\ket{\Psi_2} 
    &=\frac{1}{\sqrt{2}}\left\{
    \ket{\uparrow}_{z}
- e^{i\varphi}\ket{\downarrow}_{z}\right\}.
\end{align}
Here $\varphi$ is an equatorial angle in the Bloch sphere. The effect of this measurement is encapsulated by the probability distribution $\mathds{P}(k|x)$ with $k = 1, 2$ and $x$ is the parameter we intend to estimate. For the discrete case, the classical Fisher information is simply \cite{hall2000quantum}
\begin{align}
    \mathds{F}_{c}(x)
    = \sum^{2}_{k=1} \left(\partial^{2}_{x} \ln\left[\mathds{P}(k|x)\right]\right)\mathds{P}(k|x),
\end{align}
where $\mathds{P}(k|x)$ is the conditional probability of getting measurement result $k$, and $\partial^{2}_{x}$ denotes the double derivative with respect to the parameter \emph{x} that is to be estimated. Using the projection operators along with the final state \eqref{ch5-1}, we find that (we have set $\Theta = \chi + \omega_0 t -\varphi $ to show a more compact form)
\begin{align}
    \mathds{F}_{c}(x)
    =\frac{\left(\left(\frac{\partial{\Delta}}{\partial x} \sin\Delta + \frac{\partial{\Gamma}}{\partial x}  \cos\Delta \right) \cos\Theta - \frac{\partial{\chi}}{\partial x}\cos\Delta  \sin\Theta\right)^{2}}
    {e^{2\Gamma} - \cos^{2}\Delta \cos^{2}\Theta}.\label{ch5-9}
\end{align}
If we disregard the effect of the initial correlations, then this expression reduces to
\begin{align}
    \mathds{F}^{\text{woc}}_{c}(x)
    =\frac{\left(\frac{\partial{\Delta}}{\partial x} \sin\Delta + \frac{\partial{\Gamma}}{\partial x}  \cos\Delta \right)^{2}}
    {\sec^{2}\left(\omega_0 t -\varphi\right) e^{2\Gamma} - \cos^{2}\Delta}. \label{ch5-10}
\end{align}
We aim to maximize the classical Fisher information [see Eq.~\eqref{ch5-9}] over the angle $\varphi$. If this maximized CFI is equal to the QFI, then we have found the optimal measurement. If the effect of initial correlations is not included, then it is clear that $\varphi=\omega_0 t$ is the optimal value. In this case, the CFI reduces to the QFI, and so we have found the optimal measurement. On the other hand, if $\chi \neq 0$, we can show that for
\begin{align}
    \varphi
    = \omega_0 t + \chi- \text{tan}^{-1} \left[\frac{\partial_{x}{\chi} \cos\Delta \left( e^{2\Gamma} - \cos^{2} \Delta \right)}
    {e^{2\Gamma} \left(\partial_{x}{\Delta} \sin\Delta + \partial_{x}{\Gamma} \cos\Delta \right)} \right], \label{ch5-11}
\end{align}
the CFI reduces to the QFI. Again, this means that we have managed to find the optimal measurement. We further support these claims by plotting both the QFI and CFI while estimating environment's cutoff frequency $\omega_c$, system-environment coupling strength $G$ and the environment's temperature $T$ (see Fig.~\ref{Q3D}), where the overlap between the CFI and the QFI shows that we have successfully found the optimal measurements to be performed.

\section{Conclusion}\label{ch5-sum}

In conclusion, we have explored the use of a two-qubit probe scheme to estimate the parameters of a harmonic oscillator environment. The key idea is that both qubits of the probe interact with the common environment. This common environment induces an indirect interaction between the qubits. The dynamics of only one of these qubits then depend sensitively on this indirect interaction on top of the decoherence induced by the environment as well as the effect of the initial system-environment correlations. Consequently, we can expect our two-qubit scheme to be better at estimating the environment parameters as compared to simply using a single qubit probe. We illustrated that this is indeed  the case by evaluating the optimal quantum Fisher information. The increase in the quantum Fisher information can, in fact, be dramatic.  

\appendix
\section{Derivation of unitary operator} \label{ch5-appA}
Our total unitary time-evolution operator for the two-qubit systems can be written as as  $U(t) 
    = e^{-iH_S t} e^{-i H_E t} U_{SE}(t)$. To find the unitary time evolution operator $U_{SE}(t)$ corresponding to $H_{SE}(t)$, we use the Magnus expansion 
\begin{align}
    U_{SE}\left(t\right)
    = \text{exp}\left\{\sum_{i=1}^\infty A_i\left(t\right)\right\}. 
\end{align}
The integrals $A_1, A_2, \cdots$  are evaluated below. For simplicity, we let $\widetilde{J}_z=\sigma_z ^{\left(1\right)} + \sigma_z ^{\left(2\right)} $. We first have
\begin{align}
    A_1
    &=-i \int_0^t H_{SE} \left(t_1\right) dt_1, \nonumber
\\
    &=-i \widetilde{J}_z \sum_r \int_0^t  \left(g_{r}^{*}b_r e^{-i\omega_r t} + g_r b_r^\dagger e^{i\omega_r t}\right)dt_1, \nonumber
\\
    &=-i \widetilde{J}_z \sum_r\left\{g_{r}^{*}b_r \frac{\left(e^{-i\omega_r t}-1\right)}{-i\omega_r} + g_r b_r^\dagger \frac{\left(e^{i\omega_r t}-1\right)}{i\omega_r}\right\} , \nonumber
\\
    &=\frac{\widetilde{J}_z}{2} \sum_r\left[ \alpha_r \left(t\right)b_r^\dagger -  \alpha_r^* \left(t\right)b_r\right],
\end{align}
with
\begin{align}
    \alpha_r\left( t \right)
    =\frac{2g_r}{\omega_r}\left(1-e^{i\omega_r t} \right).
\end{align}
Now in order to calculate $A_2$, we need to determine the commutator $ \left[H_{SE} \left(t_1\right),H_{SE} \left(t_2\right)\right] $. This comes out to be 
\begin{align*}
    &\left[H_{SE} \left(t_1\right),H_{SE} \left(t_2\right)\right]\\
    &=4i\left(\mathds{1}+\sigma_z ^{\left(1\right)}  \sigma_z ^{\left(2\right)}\right) \sum_r \abs{g_r}^2 \sin\left[\omega_r (t_2-t_1)\right].
\end{align*}
Therefore
\begin{align}
    A_2
    &=-\frac{1}{2} \int _0^t dt_1\int_0^{t_1} \left[H_{SE} \left(t_1\right),H_{SE} \left(t_2\right)\right] \, dt_{2},\nonumber
\\
    &=-2i\sum_r\abs{g_r}^2\left(\mathds{1}+\sigma_z ^{\left(1\right)}  \sigma_z ^{\left(2\right)}\right)  \int _0^t dt_1\nonumber
\\    
    &\times \int_0^{t_1} \sin\left[\omega_k (t_2-t_1)\right]   \, dt_{2},\nonumber
\\
    &= -\frac{i}{2} \left(\mathds{1}+\sigma_z ^{\left(1\right)}  \sigma_z ^{\left(2\right)}\right) \sum_r\frac{4\abs{g_r}^2}{\omega_r^2}\left[\sin(\omega_r t) - \omega_r t\right],\nonumber
\\
    &= -\frac{i}{2}\left(\mathds{1}+\sigma_z ^{\left(1\right)} \sigma_z ^{\left(2\right)}\right) \Delta \left( t \right),
\end{align}
with
\begin{align}
    \Delta \left( t \right) 
    =\sum_r   \frac{4\abs{g_r}^2}{\omega_r^2}\left[\sin(\omega_r t) - \omega_r t\right].
\end{align}
This is simply a c-number, which means that $A_3=A_4=A_{5}= \cdots =0$. The expression of the exact unitary time-evolution operator can then be written as
\begin{align}
    U\left( t \right)  
    &=\text{exp}\left\{-i \left(\frac{\omega_0}{2}\left( \sigma_z ^{\left(1\right)} + \sigma_z ^{\left(2\right)}\right) + \sum_r \omega_r b_{r}^{\dagger} b_r \right)t\right\}\nonumber
\\    
    &\times \text{exp}\Bigg\{\frac{1}{2}\left( \sigma_z ^{\left(1\right)} + \sigma_z ^{\left(2\right)}\right) \sum_r\left[ \alpha_r \left(t\right)b_r^\dagger -  \alpha_r^* \left(t\right)b_r\right]\nonumber
\\    
    &-\frac{i}{2}\left(\mathds{1}+\sigma_z ^{\left(1\right)} \sigma_z ^{\left(2\right)}\right) \Delta \left( t \right)\Bigg\}.
\end{align}

\section{Dynamics with initial system-environment correlations}\label{ch5-appB}

We prepare the initial system-environment with a projective measurement described by the projection operator $\ket{\psi}\bra{\psi}$. The initial joint system-environment state can then be written as
\begin{align}
    \rho \left(0\right) 
    =\ket{\psi}\bra{\psi} \otimes \frac{\bra{\psi} e^{-\beta H} \ket{\psi}}{Z},
\end{align} 
where $Z$ is the partition function. Labeling the joint eigenstates of $\sigma_z^{(1)}$ and $\sigma_z^{(2)}$ as $\ket{p,q}$, we have 
\begin{align}
    Z
    &=  \sum_{p,q}\text{Tr}_{\text{S,E}}\Big\{ e^{-\frac{\beta\omega_0}{2}\left( \sigma_z ^{\left(1\right)} + \sigma_z ^{\left(2\right)}\right)}\nonumber   
\\    
    &\times e^{-\beta \left\{H_E+\left( \sigma_z ^{\left(1\right)} + \sigma_z ^{\left(2\right)}\right) \sum_{r} \left(g_{r}^{*}b_r  + g_r b_r^\dagger \right)\right\}}\ket{p,q}\bra{p,q} \Big\}, \nonumber 
\\
    &=\sum_{p,q}\text{Tr}_{\text{S,E}}\Big\{ e^{-\frac{\beta\omega_0}{2}\left( p+q\right)}  \nonumber
\\
   &\times e^{-\beta \left\{H_E+\left( p+q\right) \sum_{r} \left(g_{r}^{*}b_r  + g_r b_r^\dagger \right)\right\}}\ket{p,q}\bra{p,q} \Big\}, \nonumber 
\end{align}    
which further simplifies to 
\begin{align}
    Z
    &=  \sum_{p,q} e^{-\frac{\beta\omega_0}{2}\left( p+q\right)} \text{Tr}_E\left\{e^{-\beta \left\{H_E+\left( p+q\right) \sum_{r} \left(g_{r}^{*}b_r  + g_r b_r^\dagger \right)\right\}} \right\},\nonumber
\\    
    &= \sum_{p,q}e^{-\frac{\beta\omega_0}{2}\left( p+q\right)} \text{Tr}_E\left\{e^{-\beta H_E^{\left(p,q\right)}} \right\},
\end{align}
where we defined the shifted Hamiltonian
\begin{align}
    H_E^{\left(p,q\right)} 
    = H_E+\left( p+q\right) \sum_{r} \left(g_{r}^{*}b_r  + g_r b_r^\dagger \right).
\end{align}
To further simplify $H_E^{\left(p,q\right)}$, we use displaced harmonic oscillator modes
\begin{align*}
    B_{r,p,q}
    &= b_r + \frac{\left( p+q\right)g_r}{\omega_r},
\\
    B^{\dagger}_{r,p,q}
    &= b_r + \frac{\left( p+q\right)g_r}{\omega_r}.
\end{align*}
Now the trace over the environment becomes 
    $\text{Tr}_{E}\left\{ e^{-\beta H_E^{\left(p,q\right)}} \right\}
    =e^{\beta \sum_r(p+q)^2\frac{\abs{g_r}^2}{\omega_r}
    }Z_E,$
with $Z_E=\text{Tr}_{E}\left\{ e^{-\beta\sum_r  \left\{\omega_r B_{r,p,q}^\dagger B_{r,p,q}
    \right\}} \right\}$. Therefore, the final form of the partition function is
\begin{align}
    Z
    &=\sum_{p,q}e^{-\frac{\beta\omega_0}{2}\left( p+q\right)}e^{\beta \sum_r(p+q)^2\frac{\abs{g_r}^2}{\omega_r}
    }Z_E.
\end{align}
In a similar manner, we can find the time-dependent factor $R_{kl,k'l'}\left(t\right)$ [taken from Eq. \eqref{ch5-3}]. We find that
\begin{align}
    R_{kl,k'l'}\left(t\right)
    &=\sum_r \left[ \widetilde{\alpha}_r\left(t\right)B_{k,l}^\dagger -\widetilde{\alpha}^*_r \left(t\right)B_{k,l} \right] + i(p+q) \widetilde{\Phi}\left(t\right),\nonumber
\end{align}
where
\begin{align}
    \widetilde{\Phi}\left(t\right) 
    =\frac{1}{2}\left(k+l-k'-l'\right)\phi\left(t\right),\nonumber
\end{align}
with $\phi\left(t\right)
    = \sum_r \frac{4\abs{g_r}^2}{\omega_r^2}\sin\left(\omega_r t\right)$. We now simplify Eq.~\eqref{ch5-3}, starting from the initial state in Eq.~\eqref{ch5-7}, as
\begin{align}
    &\text{Tr}_{\text{S,E}} \left\{  \rho \left(0\right) e^{-R_{kl,k'l'}\left(t\right)}P_{kl,k'l'} \right\}\nonumber
\\    
    &=\frac{1}{Z}\sum_{p,q}e^{-\frac{\beta\omega_0}{2}\left( p+q\right)}Z_E \bra{p,q}\ket{\psi}\bra{\psi}  P_{kl,k'l'}\ket{\psi}\bra{\psi}\ket{p,q} \nonumber
\\    
    &\times e^{-i(p+q)\widetilde{\Phi}\left(t\right)}e^{\beta \frac{\mathcal{C}}{4}}  e^{-\frac{1}{4}\left(k+l-k'-l'\right)^2\Gamma\left(t\right)}.\nonumber
\end{align}
Using these results, the two-qubit density matrix takes the form
\begin{align}
    &\left[\rho_S \left(t\right)\right]_{k',l';k,l} 
    =\left[\rho_S \left(0\right)\right]_{k',l';k,l}X\left(t\right)\nonumber
\\    
    &\times e^{-i\frac{\omega_0}{2} \left(k'+l'-k-l\right)t} e^{-i\frac{\Delta\left(t\right)}{2} \left(k'l'-kl\right)}e^{ -\frac{1}{4}\left(k+l-k'-l'\right)^2\Gamma\left(t\right)}  \nonumber,
\end{align}
where $X\left(t\right)$ has been defined in the main text.

\bibliography{bib.bib}

\begin{thebibliography}{66}%
\makeatletter
\providecommand \@ifxundefined [1]{%
 \@ifx{#1\undefined}
}%
\providecommand \@ifnum [1]{%
 \ifnum #1\expandafter \@firstoftwo
 \else \expandafter \@secondoftwo
 \fi
}%
\providecommand \@ifx [1]{%
 \ifx #1\expandafter \@firstoftwo
 \else \expandafter \@secondoftwo
 \fi
}%
\providecommand \natexlab [1]{#1}%
\providecommand \enquote  [1]{``#1''}%
\providecommand \bibnamefont  [1]{#1}%
\providecommand \bibfnamefont [1]{#1}%
\providecommand \citenamefont [1]{#1}%
\providecommand \href@noop [0]{\@secondoftwo}%
\providecommand \href [0]{\begingroup \@sanitize@url \@href}%
\providecommand \@href[1]{\@@startlink{#1}\@@href}%
\providecommand \@@href[1]{\endgroup#1\@@endlink}%
\providecommand \@sanitize@url [0]{\catcode `\\12\catcode `\$12\catcode
  `\&12\catcode `\#12\catcode `\^12\catcode `\_12\catcode `\%12\relax}%
\providecommand \@@startlink[1]{}%
\providecommand \@@endlink[0]{}%
\providecommand \url  [0]{\begingroup\@sanitize@url \@url }%
\providecommand \@url [1]{\endgroup\@href {#1}{\urlprefix }}%
\providecommand \urlprefix  [0]{URL }%
\providecommand \Eprint [0]{\href }%
\providecommand \doibase [0]{https://doi.org/}%
\providecommand \selectlanguage [0]{\@gobble}%
\providecommand \bibinfo  [0]{\@secondoftwo}%
\providecommand \bibfield  [0]{\@secondoftwo}%
\providecommand \translation [1]{[#1]}%
\providecommand \BibitemOpen [0]{}%
\providecommand \bibitemStop [0]{}%
\providecommand \bibitemNoStop [0]{.\EOS\space}%
\providecommand \EOS [0]{\spacefactor3000\relax}%
\providecommand \BibitemShut  [1]{\csname bibitem#1\endcsname}%
\let\auto@bib@innerbib\@empty
\bibitem [{\citenamefont {Nielsen}\ and\ \citenamefont
  {Chuang}(2002)}]{nielsen2002quantum}%
  \BibitemOpen
  \bibfield  {author} {\bibinfo {author} {\bibfnamefont {M.~A.}\ \bibnamefont
  {Nielsen}}\ and\ \bibinfo {author} {\bibfnamefont {I.}~\bibnamefont
  {Chuang}},\ }\href@noop {} {\bibinfo {title} {Quantum computation and quantum
  information}} (\bibinfo {year} {2002})\BibitemShut {NoStop}%
\bibitem [{\citenamefont {Haroche}\ \emph {et~al.}(2014)\citenamefont
  {Haroche}, \citenamefont {Raimond},\ and\ \citenamefont
  {Dowling}}]{haroche2014exploring}%
  \BibitemOpen
  \bibfield  {author} {\bibinfo {author} {\bibfnamefont {S.}~\bibnamefont
  {Haroche}}, \bibinfo {author} {\bibfnamefont {J.-M.}\ \bibnamefont
  {Raimond}},\ and\ \bibinfo {author} {\bibfnamefont {J.~P.}\ \bibnamefont
  {Dowling}},\ }\bibfield  {title} {\bibinfo {title} {Exploring the quantum:
  Atoms, cavities, and photons.},\ }\href@noop {} {\bibfield  {journal}
  {\bibinfo  {journal} {American Journal of Physics}\ }\textbf {\bibinfo
  {volume} {82}},\ \bibinfo {pages} {86} (\bibinfo {year} {2014})}\BibitemShut
  {NoStop}%
\bibitem [{\citenamefont {Schlosshauer}(2007)}]{schlosshauer2007decoherence}%
  \BibitemOpen
  \bibfield  {author} {\bibinfo {author} {\bibfnamefont {M.~A.}\ \bibnamefont
  {Schlosshauer}},\ }\href@noop {} {\emph {\bibinfo {title} {Decoherence: and
  the quantum-to-classical transition}}}\ (\bibinfo  {publisher} {Springer
  Science \& Business Media},\ \bibinfo {year} {2007})\BibitemShut {NoStop}%
\bibitem [{\citenamefont {Breuer}\ \emph {et~al.}(2002)\citenamefont {Breuer},
  \citenamefont {Petruccione} \emph {et~al.}}]{breuer2002theory}%
  \BibitemOpen
  \bibfield  {author} {\bibinfo {author} {\bibfnamefont {H.-P.}\ \bibnamefont
  {Breuer}}, \bibinfo {author} {\bibfnamefont {F.}~\bibnamefont {Petruccione}},
  \emph {et~al.},\ }\href@noop {} {\emph {\bibinfo {title} {The theory of open
  quantum systems}}}\ (\bibinfo  {publisher} {Oxford University Press on
  Demand},\ \bibinfo {year} {2002})\BibitemShut {NoStop}%
\bibitem [{\citenamefont {Brunelli}\ \emph {et~al.}(2011)\citenamefont
  {Brunelli}, \citenamefont {Olivares},\ and\ \citenamefont
  {Paris}}]{brunelli2011qubit}%
  \BibitemOpen
  \bibfield  {author} {\bibinfo {author} {\bibfnamefont {M.}~\bibnamefont
  {Brunelli}}, \bibinfo {author} {\bibfnamefont {S.}~\bibnamefont {Olivares}},\
  and\ \bibinfo {author} {\bibfnamefont {M.~G.}\ \bibnamefont {Paris}},\
  }\bibfield  {title} {\bibinfo {title} {Qubit thermometry for micromechanical
  resonators},\ }\href@noop {} {\bibfield  {journal} {\bibinfo  {journal}
  {Physical Review A}\ }\textbf {\bibinfo {volume} {84}},\ \bibinfo {pages}
  {032105} (\bibinfo {year} {2011})}\BibitemShut {NoStop}%
\bibitem [{\citenamefont {Brunelli}\ \emph {et~al.}(2012)\citenamefont
  {Brunelli}, \citenamefont {Olivares}, \citenamefont {Paternostro},\ and\
  \citenamefont {Paris}}]{brunelli2012qubit}%
  \BibitemOpen
  \bibfield  {author} {\bibinfo {author} {\bibfnamefont {M.}~\bibnamefont
  {Brunelli}}, \bibinfo {author} {\bibfnamefont {S.}~\bibnamefont {Olivares}},
  \bibinfo {author} {\bibfnamefont {M.}~\bibnamefont {Paternostro}},\ and\
  \bibinfo {author} {\bibfnamefont {M.~G.}\ \bibnamefont {Paris}},\ }\bibfield
  {title} {\bibinfo {title} {Qubit-assisted thermometry of a quantum harmonic
  oscillator},\ }\href@noop {} {\bibfield  {journal} {\bibinfo  {journal}
  {Physical Review A}\ }\textbf {\bibinfo {volume} {86}},\ \bibinfo {pages}
  {012125} (\bibinfo {year} {2012})}\BibitemShut {NoStop}%
\bibitem [{\citenamefont {Neumann}\ \emph {et~al.}(2013)\citenamefont
  {Neumann}, \citenamefont {Jakobi}, \citenamefont {Dolde}, \citenamefont
  {Burk}, \citenamefont {Reuter}, \citenamefont {Waldherr}, \citenamefont
  {Honert}, \citenamefont {Wolf}, \citenamefont {Brunner}, \citenamefont {Shim}
  \emph {et~al.}}]{neumann2013high}%
  \BibitemOpen
  \bibfield  {author} {\bibinfo {author} {\bibfnamefont {P.}~\bibnamefont
  {Neumann}}, \bibinfo {author} {\bibfnamefont {I.}~\bibnamefont {Jakobi}},
  \bibinfo {author} {\bibfnamefont {F.}~\bibnamefont {Dolde}}, \bibinfo
  {author} {\bibfnamefont {C.}~\bibnamefont {Burk}}, \bibinfo {author}
  {\bibfnamefont {R.}~\bibnamefont {Reuter}}, \bibinfo {author} {\bibfnamefont
  {G.}~\bibnamefont {Waldherr}}, \bibinfo {author} {\bibfnamefont
  {J.}~\bibnamefont {Honert}}, \bibinfo {author} {\bibfnamefont
  {T.}~\bibnamefont {Wolf}}, \bibinfo {author} {\bibfnamefont {A.}~\bibnamefont
  {Brunner}}, \bibinfo {author} {\bibfnamefont {J.~H.}\ \bibnamefont {Shim}},
  \emph {et~al.},\ }\bibfield  {title} {\bibinfo {title} {High-precision
  nanoscale temperature sensing using single defects in diamond},\ }\href@noop
  {} {\bibfield  {journal} {\bibinfo  {journal} {Nano letters}\ }\textbf
  {\bibinfo {volume} {13}},\ \bibinfo {pages} {2738} (\bibinfo {year}
  {2013})}\BibitemShut {NoStop}%
\bibitem [{\citenamefont {Benedetti}\ \emph {et~al.}(2014)\citenamefont
  {Benedetti}, \citenamefont {Buscemi}, \citenamefont {Bordone},\ and\
  \citenamefont {Paris}}]{benedetti2014quantum}%
  \BibitemOpen
  \bibfield  {author} {\bibinfo {author} {\bibfnamefont {C.}~\bibnamefont
  {Benedetti}}, \bibinfo {author} {\bibfnamefont {F.}~\bibnamefont {Buscemi}},
  \bibinfo {author} {\bibfnamefont {P.}~\bibnamefont {Bordone}},\ and\ \bibinfo
  {author} {\bibfnamefont {M.~G.}\ \bibnamefont {Paris}},\ }\bibfield  {title}
  {\bibinfo {title} {Quantum probes for the spectral properties of a classical
  environment},\ }\href@noop {} {\bibfield  {journal} {\bibinfo  {journal}
  {Physical Review A}\ }\textbf {\bibinfo {volume} {89}},\ \bibinfo {pages}
  {032114} (\bibinfo {year} {2014})}\BibitemShut {NoStop}%
\bibitem [{\citenamefont {Correa}\ \emph {et~al.}(2015)\citenamefont {Correa},
  \citenamefont {Mehboudi}, \citenamefont {Adesso},\ and\ \citenamefont
  {Sanpera}}]{correa2015individual}%
  \BibitemOpen
  \bibfield  {author} {\bibinfo {author} {\bibfnamefont {L.~A.}\ \bibnamefont
  {Correa}}, \bibinfo {author} {\bibfnamefont {M.}~\bibnamefont {Mehboudi}},
  \bibinfo {author} {\bibfnamefont {G.}~\bibnamefont {Adesso}},\ and\ \bibinfo
  {author} {\bibfnamefont {A.}~\bibnamefont {Sanpera}},\ }\bibfield  {title}
  {\bibinfo {title} {Individual quantum probes for optimal thermometry},\
  }\href@noop {} {\bibfield  {journal} {\bibinfo  {journal} {Physical review
  letters}\ }\textbf {\bibinfo {volume} {114}},\ \bibinfo {pages} {220405}
  (\bibinfo {year} {2015})}\BibitemShut {NoStop}%
\bibitem [{\citenamefont {Elliott}\ and\ \citenamefont
  {Johnson}(2016)}]{elliott2016nondestructive}%
  \BibitemOpen
  \bibfield  {author} {\bibinfo {author} {\bibfnamefont {T.~J.}\ \bibnamefont
  {Elliott}}\ and\ \bibinfo {author} {\bibfnamefont {T.~H.}\ \bibnamefont
  {Johnson}},\ }\bibfield  {title} {\bibinfo {title} {Nondestructive probing of
  means, variances, and correlations of ultracold-atomic-system densities via
  qubit impurities},\ }\href@noop {} {\bibfield  {journal} {\bibinfo  {journal}
  {Physical Review A}\ }\textbf {\bibinfo {volume} {93}},\ \bibinfo {pages}
  {043612} (\bibinfo {year} {2016})}\BibitemShut {NoStop}%
\bibitem [{\citenamefont {Norris}\ \emph {et~al.}(2016)\citenamefont {Norris},
  \citenamefont {Paz-Silva},\ and\ \citenamefont {Viola}}]{norris2016qubit}%
  \BibitemOpen
  \bibfield  {author} {\bibinfo {author} {\bibfnamefont {L.~M.}\ \bibnamefont
  {Norris}}, \bibinfo {author} {\bibfnamefont {G.~A.}\ \bibnamefont
  {Paz-Silva}},\ and\ \bibinfo {author} {\bibfnamefont {L.}~\bibnamefont
  {Viola}},\ }\bibfield  {title} {\bibinfo {title} {Qubit noise spectroscopy
  for non-gaussian dephasing environments},\ }\href@noop {} {\bibfield
  {journal} {\bibinfo  {journal} {Physical Review Letters}\ }\textbf {\bibinfo
  {volume} {116}},\ \bibinfo {pages} {150503} (\bibinfo {year}
  {2016})}\BibitemShut {NoStop}%
\bibitem [{\citenamefont {Tamascelli}\ \emph {et~al.}(2016)\citenamefont
  {Tamascelli}, \citenamefont {Benedetti}, \citenamefont {Olivares},\ and\
  \citenamefont {Paris}}]{tamascelli2016characterization}%
  \BibitemOpen
  \bibfield  {author} {\bibinfo {author} {\bibfnamefont {D.}~\bibnamefont
  {Tamascelli}}, \bibinfo {author} {\bibfnamefont {C.}~\bibnamefont
  {Benedetti}}, \bibinfo {author} {\bibfnamefont {S.}~\bibnamefont
  {Olivares}},\ and\ \bibinfo {author} {\bibfnamefont {M.~G.}\ \bibnamefont
  {Paris}},\ }\bibfield  {title} {\bibinfo {title} {Characterization of qubit
  chains by feynman probes},\ }\href@noop {} {\bibfield  {journal} {\bibinfo
  {journal} {Physical Review A}\ }\textbf {\bibinfo {volume} {94}},\ \bibinfo
  {pages} {042129} (\bibinfo {year} {2016})}\BibitemShut {NoStop}%
\bibitem [{\citenamefont {Streif}\ \emph {et~al.}(2016)\citenamefont {Streif},
  \citenamefont {Buchleitner}, \citenamefont {Jaksch},\ and\ \citenamefont
  {Mur-Petit}}]{streif2016measuring}%
  \BibitemOpen
  \bibfield  {author} {\bibinfo {author} {\bibfnamefont {M.}~\bibnamefont
  {Streif}}, \bibinfo {author} {\bibfnamefont {A.}~\bibnamefont {Buchleitner}},
  \bibinfo {author} {\bibfnamefont {D.}~\bibnamefont {Jaksch}},\ and\ \bibinfo
  {author} {\bibfnamefont {J.}~\bibnamefont {Mur-Petit}},\ }\bibfield  {title}
  {\bibinfo {title} {Measuring correlations of cold-atom systems using multiple
  quantum probes},\ }\href@noop {} {\bibfield  {journal} {\bibinfo  {journal}
  {Physical Review A}\ }\textbf {\bibinfo {volume} {94}},\ \bibinfo {pages}
  {053634} (\bibinfo {year} {2016})}\BibitemShut {NoStop}%
\bibitem [{\citenamefont {Benedetti}\ \emph {et~al.}(2018)\citenamefont
  {Benedetti}, \citenamefont {Sehdaran}, \citenamefont {Zandi},\ and\
  \citenamefont {Paris}}]{benedetti2018quantum}%
  \BibitemOpen
  \bibfield  {author} {\bibinfo {author} {\bibfnamefont {C.}~\bibnamefont
  {Benedetti}}, \bibinfo {author} {\bibfnamefont {F.~S.}\ \bibnamefont
  {Sehdaran}}, \bibinfo {author} {\bibfnamefont {M.~H.}\ \bibnamefont
  {Zandi}},\ and\ \bibinfo {author} {\bibfnamefont {M.~G.}\ \bibnamefont
  {Paris}},\ }\bibfield  {title} {\bibinfo {title} {Quantum probes for the
  cutoff frequency of ohmic environments},\ }\href@noop {} {\bibfield
  {journal} {\bibinfo  {journal} {Physical Review A}\ }\textbf {\bibinfo
  {volume} {97}},\ \bibinfo {pages} {012126} (\bibinfo {year}
  {2018})}\BibitemShut {NoStop}%
\bibitem [{\citenamefont {Cosco}\ \emph {et~al.}(2017)\citenamefont {Cosco},
  \citenamefont {Borrelli}, \citenamefont {Plastina},\ and\ \citenamefont
  {Maniscalco}}]{cosco2017momentum}%
  \BibitemOpen
  \bibfield  {author} {\bibinfo {author} {\bibfnamefont {F.}~\bibnamefont
  {Cosco}}, \bibinfo {author} {\bibfnamefont {M.}~\bibnamefont {Borrelli}},
  \bibinfo {author} {\bibfnamefont {F.}~\bibnamefont {Plastina}},\ and\
  \bibinfo {author} {\bibfnamefont {S.}~\bibnamefont {Maniscalco}},\ }\bibfield
   {title} {\bibinfo {title} {Momentum-resolved and correlation spectroscopy
  using quantum probes},\ }\href@noop {} {\bibfield  {journal} {\bibinfo
  {journal} {Physical Review A}\ }\textbf {\bibinfo {volume} {95}},\ \bibinfo
  {pages} {053620} (\bibinfo {year} {2017})}\BibitemShut {NoStop}%
\bibitem [{\citenamefont {Sone}\ and\ \citenamefont
  {Cappellaro}(2017)}]{sone2017exact}%
  \BibitemOpen
  \bibfield  {author} {\bibinfo {author} {\bibfnamefont {A.}~\bibnamefont
  {Sone}}\ and\ \bibinfo {author} {\bibfnamefont {P.}~\bibnamefont
  {Cappellaro}},\ }\bibfield  {title} {\bibinfo {title} {Exact dimension
  estimation of interacting qubit systems assisted by a single quantum probe},\
  }\href@noop {} {\bibfield  {journal} {\bibinfo  {journal} {Physical Review
  A}\ }\textbf {\bibinfo {volume} {96}},\ \bibinfo {pages} {062334} (\bibinfo
  {year} {2017})}\BibitemShut {NoStop}%
\bibitem [{\citenamefont {Salari~Sehdaran}\ \emph {et~al.}(2019)\citenamefont
  {Salari~Sehdaran}, \citenamefont {Bina}, \citenamefont {Benedetti},\ and\
  \citenamefont {Paris}}]{salari2019quantum}%
  \BibitemOpen
  \bibfield  {author} {\bibinfo {author} {\bibfnamefont {F.}~\bibnamefont
  {Salari~Sehdaran}}, \bibinfo {author} {\bibfnamefont {M.}~\bibnamefont
  {Bina}}, \bibinfo {author} {\bibfnamefont {C.}~\bibnamefont {Benedetti}},\
  and\ \bibinfo {author} {\bibfnamefont {M.~G.}\ \bibnamefont {Paris}},\
  }\bibfield  {title} {\bibinfo {title} {Quantum probes for ohmic environments
  at thermal equilibrium},\ }\href@noop {} {\bibfield  {journal} {\bibinfo
  {journal} {Entropy}\ }\textbf {\bibinfo {volume} {21}},\ \bibinfo {pages}
  {486} (\bibinfo {year} {2019})}\BibitemShut {NoStop}%
\bibitem [{\citenamefont {Razavian}\ \emph {et~al.}(2019)\citenamefont
  {Razavian}, \citenamefont {Benedetti}, \citenamefont {Bina}, \citenamefont
  {Akbari-Kourbolagh},\ and\ \citenamefont {Paris}}]{razavian2019quantum}%
  \BibitemOpen
  \bibfield  {author} {\bibinfo {author} {\bibfnamefont {S.}~\bibnamefont
  {Razavian}}, \bibinfo {author} {\bibfnamefont {C.}~\bibnamefont {Benedetti}},
  \bibinfo {author} {\bibfnamefont {M.}~\bibnamefont {Bina}}, \bibinfo {author}
  {\bibfnamefont {Y.}~\bibnamefont {Akbari-Kourbolagh}},\ and\ \bibinfo
  {author} {\bibfnamefont {M.~G.}\ \bibnamefont {Paris}},\ }\bibfield  {title}
  {\bibinfo {title} {Quantum thermometry by single-qubit dephasing},\
  }\href@noop {} {\bibfield  {journal} {\bibinfo  {journal} {The European
  Physical Journal Plus}\ }\textbf {\bibinfo {volume} {134}},\ \bibinfo {pages}
  {284} (\bibinfo {year} {2019})}\BibitemShut {NoStop}%
\bibitem [{\citenamefont {Gebbia}\ \emph {et~al.}(2020)\citenamefont {Gebbia},
  \citenamefont {Benedetti}, \citenamefont {Benatti}, \citenamefont
  {Floreanini}, \citenamefont {Bina},\ and\ \citenamefont
  {Paris}}]{gebbia2020two}%
  \BibitemOpen
  \bibfield  {author} {\bibinfo {author} {\bibfnamefont {F.}~\bibnamefont
  {Gebbia}}, \bibinfo {author} {\bibfnamefont {C.}~\bibnamefont {Benedetti}},
  \bibinfo {author} {\bibfnamefont {F.}~\bibnamefont {Benatti}}, \bibinfo
  {author} {\bibfnamefont {R.}~\bibnamefont {Floreanini}}, \bibinfo {author}
  {\bibfnamefont {M.}~\bibnamefont {Bina}},\ and\ \bibinfo {author}
  {\bibfnamefont {M.~G.}\ \bibnamefont {Paris}},\ }\bibfield  {title} {\bibinfo
  {title} {Two-qubit quantum probes for the temperature of an ohmic
  environment},\ }\href@noop {} {\bibfield  {journal} {\bibinfo  {journal}
  {Physical Review A}\ }\textbf {\bibinfo {volume} {101}},\ \bibinfo {pages}
  {032112} (\bibinfo {year} {2020})}\BibitemShut {NoStop}%
\bibitem [{\citenamefont {Wu}\ and\ \citenamefont {Shi}(2020)}]{wu2020quantum}%
  \BibitemOpen
  \bibfield  {author} {\bibinfo {author} {\bibfnamefont {W.}~\bibnamefont
  {Wu}}\ and\ \bibinfo {author} {\bibfnamefont {C.}~\bibnamefont {Shi}},\
  }\bibfield  {title} {\bibinfo {title} {Quantum parameter estimation in a
  dissipative environment},\ }\href@noop {} {\bibfield  {journal} {\bibinfo
  {journal} {Physical Review A}\ }\textbf {\bibinfo {volume} {102}},\ \bibinfo
  {pages} {032607} (\bibinfo {year} {2020})}\BibitemShut {NoStop}%
\bibitem [{\citenamefont {Tamascelli}\ \emph {et~al.}(2020)\citenamefont
  {Tamascelli}, \citenamefont {Benedetti}, \citenamefont {Breuer},\ and\
  \citenamefont {Paris}}]{tamascelli2020quantum}%
  \BibitemOpen
  \bibfield  {author} {\bibinfo {author} {\bibfnamefont {D.}~\bibnamefont
  {Tamascelli}}, \bibinfo {author} {\bibfnamefont {C.}~\bibnamefont
  {Benedetti}}, \bibinfo {author} {\bibfnamefont {H.-P.}\ \bibnamefont
  {Breuer}},\ and\ \bibinfo {author} {\bibfnamefont {M.~G.}\ \bibnamefont
  {Paris}},\ }\bibfield  {title} {\bibinfo {title} {Quantum probing beyond pure
  dephasing},\ }\href@noop {} {\bibfield  {journal} {\bibinfo  {journal} {New
  Journal of Physics}\ }\textbf {\bibinfo {volume} {22}},\ \bibinfo {pages}
  {083027} (\bibinfo {year} {2020})}\BibitemShut {NoStop}%
\bibitem [{\citenamefont {Gianani}\ \emph {et~al.}(2020)\citenamefont
  {Gianani}, \citenamefont {Farina}, \citenamefont {Barbieri}, \citenamefont
  {Cimini}, \citenamefont {Cavina},\ and\ \citenamefont
  {Giovannetti}}]{gianani2020discrimination}%
  \BibitemOpen
  \bibfield  {author} {\bibinfo {author} {\bibfnamefont {I.}~\bibnamefont
  {Gianani}}, \bibinfo {author} {\bibfnamefont {D.}~\bibnamefont {Farina}},
  \bibinfo {author} {\bibfnamefont {M.}~\bibnamefont {Barbieri}}, \bibinfo
  {author} {\bibfnamefont {V.}~\bibnamefont {Cimini}}, \bibinfo {author}
  {\bibfnamefont {V.}~\bibnamefont {Cavina}},\ and\ \bibinfo {author}
  {\bibfnamefont {V.}~\bibnamefont {Giovannetti}},\ }\bibfield  {title}
  {\bibinfo {title} {Discrimination of thermal baths by single-qubit probes},\
  }\href@noop {} {\bibfield  {journal} {\bibinfo  {journal} {Physical Review
  Research}\ }\textbf {\bibinfo {volume} {2}},\ \bibinfo {pages} {033497}
  (\bibinfo {year} {2020})}\BibitemShut {NoStop}%
\bibitem [{\citenamefont {Helstrom}(1976)}]{helstrom1976quantum}%
  \BibitemOpen
  \bibfield  {author} {\bibinfo {author} {\bibfnamefont {C.}~\bibnamefont
  {Helstrom}},\ }\bibfield  {title} {\bibinfo {title} {Quantum detection and
  estimation theory, ser},\ }\href@noop {} {\bibfield  {journal} {\bibinfo
  {journal} {Mathematics in Science and Engineering. New York: Academic Press}\
  }\textbf {\bibinfo {volume} {123}} (\bibinfo {year} {1976})}\BibitemShut
  {NoStop}%
\bibitem [{\citenamefont {Fujiwara}(2001)}]{fujiwara2001quantum}%
  \BibitemOpen
  \bibfield  {author} {\bibinfo {author} {\bibfnamefont {A.}~\bibnamefont
  {Fujiwara}},\ }\bibfield  {title} {\bibinfo {title} {Quantum channel
  identification problem},\ }\href@noop {} {\bibfield  {journal} {\bibinfo
  {journal} {Physical Review A}\ }\textbf {\bibinfo {volume} {63}},\ \bibinfo
  {pages} {042304} (\bibinfo {year} {2001})}\BibitemShut {NoStop}%
\bibitem [{\citenamefont {Monras}(2006)}]{monras2006optimal}%
  \BibitemOpen
  \bibfield  {author} {\bibinfo {author} {\bibfnamefont {A.}~\bibnamefont
  {Monras}},\ }\bibfield  {title} {\bibinfo {title} {Optimal phase measurements
  with pure gaussian states},\ }\href@noop {} {\bibfield  {journal} {\bibinfo
  {journal} {Physical Review A}\ }\textbf {\bibinfo {volume} {73}},\ \bibinfo
  {pages} {033821} (\bibinfo {year} {2006})}\BibitemShut {NoStop}%
\bibitem [{\citenamefont {Paris}(2009)}]{paris2009quantum}%
  \BibitemOpen
  \bibfield  {author} {\bibinfo {author} {\bibfnamefont {M.~G.}\ \bibnamefont
  {Paris}},\ }\bibfield  {title} {\bibinfo {title} {Quantum estimation for
  quantum technology},\ }\href@noop {} {\bibfield  {journal} {\bibinfo
  {journal} {International Journal of Quantum Information}\ }\textbf {\bibinfo
  {volume} {7}},\ \bibinfo {pages} {125} (\bibinfo {year} {2009})}\BibitemShut
  {NoStop}%
\bibitem [{\citenamefont {Monras}\ and\ \citenamefont
  {Paris}(2007)}]{monras2007optimal}%
  \BibitemOpen
  \bibfield  {author} {\bibinfo {author} {\bibfnamefont {A.}~\bibnamefont
  {Monras}}\ and\ \bibinfo {author} {\bibfnamefont {M.~G.}\ \bibnamefont
  {Paris}},\ }\bibfield  {title} {\bibinfo {title} {Optimal quantum estimation
  of loss in bosonic channels},\ }\href@noop {} {\bibfield  {journal} {\bibinfo
   {journal} {Physical review letters}\ }\textbf {\bibinfo {volume} {98}},\
  \bibinfo {pages} {160401} (\bibinfo {year} {2007})}\BibitemShut {NoStop}%
\bibitem [{\citenamefont {Genoni}\ \emph {et~al.}(2011)\citenamefont {Genoni},
  \citenamefont {Olivares},\ and\ \citenamefont {Paris}}]{genoni2011optical}%
  \BibitemOpen
  \bibfield  {author} {\bibinfo {author} {\bibfnamefont {M.~G.}\ \bibnamefont
  {Genoni}}, \bibinfo {author} {\bibfnamefont {S.}~\bibnamefont {Olivares}},\
  and\ \bibinfo {author} {\bibfnamefont {M.~G.}\ \bibnamefont {Paris}},\
  }\bibfield  {title} {\bibinfo {title} {Optical phase estimation in the
  presence of phase diffusion},\ }\href@noop {} {\bibfield  {journal} {\bibinfo
   {journal} {Physical review letters}\ }\textbf {\bibinfo {volume} {106}},\
  \bibinfo {pages} {153603} (\bibinfo {year} {2011})}\BibitemShut {NoStop}%
\bibitem [{\citenamefont {Spagnolo}\ \emph {et~al.}(2012)\citenamefont
  {Spagnolo}, \citenamefont {Vitelli}, \citenamefont {Lucivero}, \citenamefont
  {Giovannetti}, \citenamefont {Maccone},\ and\ \citenamefont
  {Sciarrino}}]{spagnolo2012phase}%
  \BibitemOpen
  \bibfield  {author} {\bibinfo {author} {\bibfnamefont {N.}~\bibnamefont
  {Spagnolo}}, \bibinfo {author} {\bibfnamefont {C.}~\bibnamefont {Vitelli}},
  \bibinfo {author} {\bibfnamefont {V.~G.}\ \bibnamefont {Lucivero}}, \bibinfo
  {author} {\bibfnamefont {V.}~\bibnamefont {Giovannetti}}, \bibinfo {author}
  {\bibfnamefont {L.}~\bibnamefont {Maccone}},\ and\ \bibinfo {author}
  {\bibfnamefont {F.}~\bibnamefont {Sciarrino}},\ }\bibfield  {title} {\bibinfo
  {title} {Phase estimation via quantum interferometry for noisy detectors},\
  }\href@noop {} {\bibfield  {journal} {\bibinfo  {journal} {Physical Review
  Letters}\ }\textbf {\bibinfo {volume} {108}},\ \bibinfo {pages} {233602}
  (\bibinfo {year} {2012})}\BibitemShut {NoStop}%
\bibitem [{\citenamefont {Pinel}\ \emph {et~al.}(2013)\citenamefont {Pinel},
  \citenamefont {Jian}, \citenamefont {Treps}, \citenamefont {Fabre},\ and\
  \citenamefont {Braun}}]{pinel2013quantum}%
  \BibitemOpen
  \bibfield  {author} {\bibinfo {author} {\bibfnamefont {O.}~\bibnamefont
  {Pinel}}, \bibinfo {author} {\bibfnamefont {P.}~\bibnamefont {Jian}},
  \bibinfo {author} {\bibfnamefont {N.}~\bibnamefont {Treps}}, \bibinfo
  {author} {\bibfnamefont {C.}~\bibnamefont {Fabre}},\ and\ \bibinfo {author}
  {\bibfnamefont {D.}~\bibnamefont {Braun}},\ }\bibfield  {title} {\bibinfo
  {title} {Quantum parameter estimation using general single-mode gaussian
  states},\ }\href@noop {} {\bibfield  {journal} {\bibinfo  {journal} {Physical
  Review A}\ }\textbf {\bibinfo {volume} {88}},\ \bibinfo {pages} {040102}
  (\bibinfo {year} {2013})}\BibitemShut {NoStop}%
\bibitem [{\citenamefont {Chaudhry}(2014)}]{chaudhry2014utilizing}%
  \BibitemOpen
  \bibfield  {author} {\bibinfo {author} {\bibfnamefont {A.~Z.}\ \bibnamefont
  {Chaudhry}},\ }\bibfield  {title} {\bibinfo {title} {Utilizing
  nitrogen-vacancy centers to measure oscillating magnetic fields},\
  }\href@noop {} {\bibfield  {journal} {\bibinfo  {journal} {Physical Review
  A}\ }\textbf {\bibinfo {volume} {90}},\ \bibinfo {pages} {042104} (\bibinfo
  {year} {2014})}\BibitemShut {NoStop}%
\bibitem [{\citenamefont {Chaudhry}(2015)}]{chaudhry2015detecting}%
  \BibitemOpen
  \bibfield  {author} {\bibinfo {author} {\bibfnamefont {A.~Z.}\ \bibnamefont
  {Chaudhry}},\ }\bibfield  {title} {\bibinfo {title} {Detecting the presence
  of weak magnetic fields using nitrogen-vacancy centers},\ }\href@noop {}
  {\bibfield  {journal} {\bibinfo  {journal} {Physical Review A}\ }\textbf
  {\bibinfo {volume} {91}},\ \bibinfo {pages} {062111} (\bibinfo {year}
  {2015})}\BibitemShut {NoStop}%
\bibitem [{\citenamefont {Benedetti}\ and\ \citenamefont
  {Paris}(2014)}]{benedetti2014characterization}%
  \BibitemOpen
  \bibfield  {author} {\bibinfo {author} {\bibfnamefont {C.}~\bibnamefont
  {Benedetti}}\ and\ \bibinfo {author} {\bibfnamefont {M.~G.}\ \bibnamefont
  {Paris}},\ }\bibfield  {title} {\bibinfo {title} {Characterization of
  classical gaussian processes using quantum probes},\ }\href@noop {}
  {\bibfield  {journal} {\bibinfo  {journal} {Physics Letters A}\ }\textbf
  {\bibinfo {volume} {378}},\ \bibinfo {pages} {2495} (\bibinfo {year}
  {2014})}\BibitemShut {NoStop}%
\bibitem [{\citenamefont {Hakim}\ and\ \citenamefont
  {Ambegaokar}(1985)}]{hakim1985quantum}%
  \BibitemOpen
  \bibfield  {author} {\bibinfo {author} {\bibfnamefont {V.}~\bibnamefont
  {Hakim}}\ and\ \bibinfo {author} {\bibfnamefont {V.}~\bibnamefont
  {Ambegaokar}},\ }\bibfield  {title} {\bibinfo {title} {Quantum theory of a
  free particle interacting with a linearly dissipative environment},\
  }\href@noop {} {\bibfield  {journal} {\bibinfo  {journal} {Physical Review
  A}\ }\textbf {\bibinfo {volume} {32}},\ \bibinfo {pages} {423} (\bibinfo
  {year} {1985})}\BibitemShut {NoStop}%
\bibitem [{\citenamefont {Haake}\ and\ \citenamefont
  {Reibold}(1985)}]{haake1985strong}%
  \BibitemOpen
  \bibfield  {author} {\bibinfo {author} {\bibfnamefont {F.}~\bibnamefont
  {Haake}}\ and\ \bibinfo {author} {\bibfnamefont {R.}~\bibnamefont
  {Reibold}},\ }\bibfield  {title} {\bibinfo {title} {Strong damping and
  low-temperature anomalies for the harmonic oscillator},\ }\href@noop {}
  {\bibfield  {journal} {\bibinfo  {journal} {Physical Review A}\ }\textbf
  {\bibinfo {volume} {32}},\ \bibinfo {pages} {2462} (\bibinfo {year}
  {1985})}\BibitemShut {NoStop}%
\bibitem [{\citenamefont {Grabert}\ \emph {et~al.}(1988)\citenamefont
  {Grabert}, \citenamefont {Schramm},\ and\ \citenamefont
  {Ingold}}]{grabert1988quantum}%
  \BibitemOpen
  \bibfield  {author} {\bibinfo {author} {\bibfnamefont {H.}~\bibnamefont
  {Grabert}}, \bibinfo {author} {\bibfnamefont {P.}~\bibnamefont {Schramm}},\
  and\ \bibinfo {author} {\bibfnamefont {G.-L.}\ \bibnamefont {Ingold}},\
  }\bibfield  {title} {\bibinfo {title} {Quantum brownian motion: The
  functional integral approach},\ }\href@noop {} {\bibfield  {journal}
  {\bibinfo  {journal} {Physics reports}\ }\textbf {\bibinfo {volume} {168}},\
  \bibinfo {pages} {115} (\bibinfo {year} {1988})}\BibitemShut {NoStop}%
\bibitem [{\citenamefont {Smith}\ and\ \citenamefont
  {Caldeira}(1990)}]{smith1990application}%
  \BibitemOpen
  \bibfield  {author} {\bibinfo {author} {\bibfnamefont {C.~M.}\ \bibnamefont
  {Smith}}\ and\ \bibinfo {author} {\bibfnamefont {A.}~\bibnamefont
  {Caldeira}},\ }\bibfield  {title} {\bibinfo {title} {Application of the
  generalized feynman-vernon approach to a simple system: The damped harmonic
  oscillator},\ }\href@noop {} {\bibfield  {journal} {\bibinfo  {journal}
  {Physical Review A}\ }\textbf {\bibinfo {volume} {41}},\ \bibinfo {pages}
  {3103} (\bibinfo {year} {1990})}\BibitemShut {NoStop}%
\bibitem [{\citenamefont {Karrlein}\ and\ \citenamefont
  {Grabert}(1997)}]{karrlein1997exact}%
  \BibitemOpen
  \bibfield  {author} {\bibinfo {author} {\bibfnamefont {R.}~\bibnamefont
  {Karrlein}}\ and\ \bibinfo {author} {\bibfnamefont {H.}~\bibnamefont
  {Grabert}},\ }\bibfield  {title} {\bibinfo {title} {Exact time evolution and
  master equations for the damped harmonic oscillator},\ }\href@noop {}
  {\bibfield  {journal} {\bibinfo  {journal} {Physical Review E}\ }\textbf
  {\bibinfo {volume} {55}},\ \bibinfo {pages} {153} (\bibinfo {year}
  {1997})}\BibitemShut {NoStop}%
\bibitem [{\citenamefont {Romero}\ and\ \citenamefont
  {Paz}(1997)}]{romero1997decoherence}%
  \BibitemOpen
  \bibfield  {author} {\bibinfo {author} {\bibfnamefont {L.~D.}\ \bibnamefont
  {Romero}}\ and\ \bibinfo {author} {\bibfnamefont {J.~P.}\ \bibnamefont
  {Paz}},\ }\bibfield  {title} {\bibinfo {title} {Decoherence and initial
  correlations in quantum brownian motion},\ }\href@noop {} {\bibfield
  {journal} {\bibinfo  {journal} {Physical Review A}\ }\textbf {\bibinfo
  {volume} {55}},\ \bibinfo {pages} {4070} (\bibinfo {year}
  {1997})}\BibitemShut {NoStop}%
\bibitem [{\citenamefont {Lutz}(2003)}]{lutz2003effect}%
  \BibitemOpen
  \bibfield  {author} {\bibinfo {author} {\bibfnamefont {E.}~\bibnamefont
  {Lutz}},\ }\bibfield  {title} {\bibinfo {title} {Effect of initial
  correlations on short-time decoherence},\ }\href@noop {} {\bibfield
  {journal} {\bibinfo  {journal} {Physical Review A}\ }\textbf {\bibinfo
  {volume} {67}},\ \bibinfo {pages} {022109} (\bibinfo {year}
  {2003})}\BibitemShut {NoStop}%
\bibitem [{\citenamefont {Banerjee}\ and\ \citenamefont
  {Ghosh}(2003)}]{banerjee2003general}%
  \BibitemOpen
  \bibfield  {author} {\bibinfo {author} {\bibfnamefont {S.}~\bibnamefont
  {Banerjee}}\ and\ \bibinfo {author} {\bibfnamefont {R.}~\bibnamefont
  {Ghosh}},\ }\bibfield  {title} {\bibinfo {title} {General quantum brownian
  motion with initially correlated and nonlinearly coupled environment},\
  }\href@noop {} {\bibfield  {journal} {\bibinfo  {journal} {Physical Review
  E}\ }\textbf {\bibinfo {volume} {67}},\ \bibinfo {pages} {056120} (\bibinfo
  {year} {2003})}\BibitemShut {NoStop}%
\bibitem [{\citenamefont {Van~Kampen}(2004)}]{van2004new}%
  \BibitemOpen
  \bibfield  {author} {\bibinfo {author} {\bibfnamefont {N.}~\bibnamefont
  {Van~Kampen}},\ }\bibfield  {title} {\bibinfo {title} {A new approach to
  noise in quantum mechanics},\ }\href@noop {} {\bibfield  {journal} {\bibinfo
  {journal} {Journal of statistical physics}\ }\textbf {\bibinfo {volume}
  {115}},\ \bibinfo {pages} {1057} (\bibinfo {year} {2004})}\BibitemShut
  {NoStop}%
\bibitem [{\citenamefont {Ban}(2009)}]{ban2009quantum}%
  \BibitemOpen
  \bibfield  {author} {\bibinfo {author} {\bibfnamefont {M.}~\bibnamefont
  {Ban}},\ }\bibfield  {title} {\bibinfo {title} {Quantum master equation for
  dephasing of a two-level system with an initial correlation},\ }\href@noop {}
  {\bibfield  {journal} {\bibinfo  {journal} {Physical Review A}\ }\textbf
  {\bibinfo {volume} {80}},\ \bibinfo {pages} {064103} (\bibinfo {year}
  {2009})}\BibitemShut {NoStop}%
\bibitem [{\citenamefont {Campisi}\ \emph {et~al.}(2009)\citenamefont
  {Campisi}, \citenamefont {Talkner},\ and\ \citenamefont
  {H{\"a}nggi}}]{campisi2009fluctuation}%
  \BibitemOpen
  \bibfield  {author} {\bibinfo {author} {\bibfnamefont {M.}~\bibnamefont
  {Campisi}}, \bibinfo {author} {\bibfnamefont {P.}~\bibnamefont {Talkner}},\
  and\ \bibinfo {author} {\bibfnamefont {P.}~\bibnamefont {H{\"a}nggi}},\
  }\bibfield  {title} {\bibinfo {title} {Fluctuation theorem for arbitrary open
  quantum systems},\ }\href@noop {} {\bibfield  {journal} {\bibinfo  {journal}
  {Physical review letters}\ }\textbf {\bibinfo {volume} {102}},\ \bibinfo
  {pages} {210401} (\bibinfo {year} {2009})}\BibitemShut {NoStop}%
\bibitem [{\citenamefont {Uchiyama}\ and\ \citenamefont
  {Aihara}(2010)}]{uchiyama2010role}%
  \BibitemOpen
  \bibfield  {author} {\bibinfo {author} {\bibfnamefont {C.}~\bibnamefont
  {Uchiyama}}\ and\ \bibinfo {author} {\bibfnamefont {M.}~\bibnamefont
  {Aihara}},\ }\bibfield  {title} {\bibinfo {title} {Role of initial quantum
  correlation in transient linear response},\ }\href@noop {} {\bibfield
  {journal} {\bibinfo  {journal} {Physical Review A}\ }\textbf {\bibinfo
  {volume} {82}},\ \bibinfo {pages} {044104} (\bibinfo {year}
  {2010})}\BibitemShut {NoStop}%
\bibitem [{\citenamefont {Dijkstra}\ and\ \citenamefont
  {Tanimura}(2010)}]{dijkstra2010non}%
  \BibitemOpen
  \bibfield  {author} {\bibinfo {author} {\bibfnamefont {A.~G.}\ \bibnamefont
  {Dijkstra}}\ and\ \bibinfo {author} {\bibfnamefont {Y.}~\bibnamefont
  {Tanimura}},\ }\bibfield  {title} {\bibinfo {title} {Non-markovian
  entanglement dynamics in the presence of system-bath coherence},\ }\href@noop
  {} {\bibfield  {journal} {\bibinfo  {journal} {Physical review letters}\
  }\textbf {\bibinfo {volume} {104}},\ \bibinfo {pages} {250401} (\bibinfo
  {year} {2010})}\BibitemShut {NoStop}%
\bibitem [{\citenamefont {Smirne}\ \emph {et~al.}(2010)\citenamefont {Smirne},
  \citenamefont {Breuer}, \citenamefont {Piilo},\ and\ \citenamefont
  {Vacchini}}]{smirne2010initial}%
  \BibitemOpen
  \bibfield  {author} {\bibinfo {author} {\bibfnamefont {A.}~\bibnamefont
  {Smirne}}, \bibinfo {author} {\bibfnamefont {H.-P.}\ \bibnamefont {Breuer}},
  \bibinfo {author} {\bibfnamefont {J.}~\bibnamefont {Piilo}},\ and\ \bibinfo
  {author} {\bibfnamefont {B.}~\bibnamefont {Vacchini}},\ }\bibfield  {title}
  {\bibinfo {title} {Initial correlations in open-systems dynamics: The
  jaynes-cummings model},\ }\href@noop {} {\bibfield  {journal} {\bibinfo
  {journal} {Physical Review A}\ }\textbf {\bibinfo {volume} {82}},\ \bibinfo
  {pages} {062114} (\bibinfo {year} {2010})}\BibitemShut {NoStop}%
\bibitem [{\citenamefont {Dajka}\ and\ \citenamefont
  {{\L}uczka}(2010)}]{dajka2010distance}%
  \BibitemOpen
  \bibfield  {author} {\bibinfo {author} {\bibfnamefont {J.}~\bibnamefont
  {Dajka}}\ and\ \bibinfo {author} {\bibfnamefont {J.}~\bibnamefont
  {{\L}uczka}},\ }\bibfield  {title} {\bibinfo {title} {Distance growth of
  quantum states due to initial system-environment correlations},\ }\href@noop
  {} {\bibfield  {journal} {\bibinfo  {journal} {Physical Review A}\ }\textbf
  {\bibinfo {volume} {82}},\ \bibinfo {pages} {012341} (\bibinfo {year}
  {2010})}\BibitemShut {NoStop}%
\bibitem [{\citenamefont {Zhang}\ \emph {et~al.}(2010)\citenamefont {Zhang},
  \citenamefont {Zou}, \citenamefont {Xia},\ and\ \citenamefont
  {Guo}}]{zhang2010different}%
  \BibitemOpen
  \bibfield  {author} {\bibinfo {author} {\bibfnamefont {Y.-J.}\ \bibnamefont
  {Zhang}}, \bibinfo {author} {\bibfnamefont {X.-B.}\ \bibnamefont {Zou}},
  \bibinfo {author} {\bibfnamefont {Y.-J.}\ \bibnamefont {Xia}},\ and\ \bibinfo
  {author} {\bibfnamefont {G.-C.}\ \bibnamefont {Guo}},\ }\bibfield  {title}
  {\bibinfo {title} {Different entanglement dynamical behaviors due to initial
  system-environment correlations},\ }\href@noop {} {\bibfield  {journal}
  {\bibinfo  {journal} {Physical Review A}\ }\textbf {\bibinfo {volume} {82}},\
  \bibinfo {pages} {022108} (\bibinfo {year} {2010})}\BibitemShut {NoStop}%
\bibitem [{\citenamefont {Tan}\ and\ \citenamefont {Zhang}(2011)}]{tan2011non}%
  \BibitemOpen
  \bibfield  {author} {\bibinfo {author} {\bibfnamefont {H.-T.}\ \bibnamefont
  {Tan}}\ and\ \bibinfo {author} {\bibfnamefont {W.-M.}\ \bibnamefont
  {Zhang}},\ }\bibfield  {title} {\bibinfo {title} {Non-markovian dynamics of
  an open quantum system with initial system-reservoir correlations: A
  nanocavity coupled to a coupled-resonator optical waveguide},\ }\href@noop {}
  {\bibfield  {journal} {\bibinfo  {journal} {Physical Review A}\ }\textbf
  {\bibinfo {volume} {83}},\ \bibinfo {pages} {032102} (\bibinfo {year}
  {2011})}\BibitemShut {NoStop}%
\bibitem [{\citenamefont {Lee}\ \emph {et~al.}(2012)\citenamefont {Lee},
  \citenamefont {Cao},\ and\ \citenamefont {Gong}}]{lee2012noncanonical}%
  \BibitemOpen
  \bibfield  {author} {\bibinfo {author} {\bibfnamefont {C.~K.}\ \bibnamefont
  {Lee}}, \bibinfo {author} {\bibfnamefont {J.}~\bibnamefont {Cao}},\ and\
  \bibinfo {author} {\bibfnamefont {J.}~\bibnamefont {Gong}},\ }\bibfield
  {title} {\bibinfo {title} {Noncanonical statistics of a spin-boson model:
  Theory and exact monte carlo simulations},\ }\href@noop {} {\bibfield
  {journal} {\bibinfo  {journal} {Physical Review E}\ }\textbf {\bibinfo
  {volume} {86}},\ \bibinfo {pages} {021109} (\bibinfo {year}
  {2012})}\BibitemShut {NoStop}%
\bibitem [{\citenamefont {Morozov}\ \emph {et~al.}(2012)\citenamefont
  {Morozov}, \citenamefont {Mathey},\ and\ \citenamefont
  {R{\"o}pke}}]{morozov2012decoherence}%
  \BibitemOpen
  \bibfield  {author} {\bibinfo {author} {\bibfnamefont {V.}~\bibnamefont
  {Morozov}}, \bibinfo {author} {\bibfnamefont {S.}~\bibnamefont {Mathey}},\
  and\ \bibinfo {author} {\bibfnamefont {G.}~\bibnamefont {R{\"o}pke}},\
  }\bibfield  {title} {\bibinfo {title} {Decoherence in an exactly solvable
  qubit model with initial qubit-environment correlations},\ }\href@noop {}
  {\bibfield  {journal} {\bibinfo  {journal} {Physical Review A}\ }\textbf
  {\bibinfo {volume} {85}},\ \bibinfo {pages} {022101} (\bibinfo {year}
  {2012})}\BibitemShut {NoStop}%
\bibitem [{\citenamefont {Semin}\ \emph {et~al.}(2012)\citenamefont {Semin},
  \citenamefont {Sinayskiy},\ and\ \citenamefont
  {Petruccione}}]{semin2012initial}%
  \BibitemOpen
  \bibfield  {author} {\bibinfo {author} {\bibfnamefont {V.}~\bibnamefont
  {Semin}}, \bibinfo {author} {\bibfnamefont {I.}~\bibnamefont {Sinayskiy}},\
  and\ \bibinfo {author} {\bibfnamefont {F.}~\bibnamefont {Petruccione}},\
  }\bibfield  {title} {\bibinfo {title} {Initial correlation in a system of a
  spin coupled to a spin bath through an intermediate spin},\ }\href@noop {}
  {\bibfield  {journal} {\bibinfo  {journal} {Physical Review A}\ }\textbf
  {\bibinfo {volume} {86}},\ \bibinfo {pages} {062114} (\bibinfo {year}
  {2012})}\BibitemShut {NoStop}%
\bibitem [{\citenamefont {Chaudhry}\ and\ \citenamefont
  {Gong}(2013{\natexlab{a}})}]{chaudhry2013amplification}%
  \BibitemOpen
  \bibfield  {author} {\bibinfo {author} {\bibfnamefont {A.~Z.}\ \bibnamefont
  {Chaudhry}}\ and\ \bibinfo {author} {\bibfnamefont {J.}~\bibnamefont
  {Gong}},\ }\bibfield  {title} {\bibinfo {title} {Amplification and
  suppression of system-bath-correlation effects in an open many-body system},\
  }\href@noop {} {\bibfield  {journal} {\bibinfo  {journal} {Physical Review
  A}\ }\textbf {\bibinfo {volume} {87}},\ \bibinfo {pages} {012129} (\bibinfo
  {year} {2013}{\natexlab{a}})}\BibitemShut {NoStop}%
\bibitem [{\citenamefont {Reina}\ \emph {et~al.}(2014)\citenamefont {Reina},
  \citenamefont {Susa},\ and\ \citenamefont {Fanchini}}]{reina2014extracting}%
  \BibitemOpen
  \bibfield  {author} {\bibinfo {author} {\bibfnamefont {J.~H.}\ \bibnamefont
  {Reina}}, \bibinfo {author} {\bibfnamefont {C.~E.}\ \bibnamefont {Susa}},\
  and\ \bibinfo {author} {\bibfnamefont {F.~F.}\ \bibnamefont {Fanchini}},\
  }\bibfield  {title} {\bibinfo {title} {Extracting information from
  qubit-environment correlations},\ }\href@noop {} {\bibfield  {journal}
  {\bibinfo  {journal} {Scientific Reports}\ }\textbf {\bibinfo {volume} {4}},\
  \bibinfo {pages} {1} (\bibinfo {year} {2014})}\BibitemShut {NoStop}%
\bibitem [{\citenamefont {Chaudhry}\ and\ \citenamefont
  {Gong}(2013{\natexlab{b}})}]{chaudhry2013role}%
  \BibitemOpen
  \bibfield  {author} {\bibinfo {author} {\bibfnamefont {A.~Z.}\ \bibnamefont
  {Chaudhry}}\ and\ \bibinfo {author} {\bibfnamefont {J.}~\bibnamefont
  {Gong}},\ }\bibfield  {title} {\bibinfo {title} {Role of initial
  system-environment correlations: A master equation approach},\ }\href@noop {}
  {\bibfield  {journal} {\bibinfo  {journal} {Physical Review A}\ }\textbf
  {\bibinfo {volume} {88}},\ \bibinfo {pages} {052107} (\bibinfo {year}
  {2013}{\natexlab{b}})}\BibitemShut {NoStop}%
\bibitem [{\citenamefont {Chaudhry}\ and\ \citenamefont
  {Gong}(2014)}]{chaudhry2014effect}%
  \BibitemOpen
  \bibfield  {author} {\bibinfo {author} {\bibfnamefont {A.~Z.}\ \bibnamefont
  {Chaudhry}}\ and\ \bibinfo {author} {\bibfnamefont {J.}~\bibnamefont
  {Gong}},\ }\bibfield  {title} {\bibinfo {title} {The effect of state
  preparation in a many-body system},\ }\href@noop {} {\bibfield  {journal}
  {\bibinfo  {journal} {Canadian Journal of Chemistry}\ }\textbf {\bibinfo
  {volume} {92}},\ \bibinfo {pages} {119} (\bibinfo {year} {2014})}\BibitemShut
  {NoStop}%
\bibitem [{\citenamefont {Zhang}\ \emph {et~al.}(2015)\citenamefont {Zhang},
  \citenamefont {Han}, \citenamefont {Xia}, \citenamefont {Yu},\ and\
  \citenamefont {Fan}}]{zhang2015role}%
  \BibitemOpen
  \bibfield  {author} {\bibinfo {author} {\bibfnamefont {Y.-J.}\ \bibnamefont
  {Zhang}}, \bibinfo {author} {\bibfnamefont {W.}~\bibnamefont {Han}}, \bibinfo
  {author} {\bibfnamefont {Y.-J.}\ \bibnamefont {Xia}}, \bibinfo {author}
  {\bibfnamefont {Y.-M.}\ \bibnamefont {Yu}},\ and\ \bibinfo {author}
  {\bibfnamefont {H.}~\bibnamefont {Fan}},\ }\bibfield  {title} {\bibinfo
  {title} {Role of initial system-bath correlation on coherence trapping},\
  }\href@noop {} {\bibfield  {journal} {\bibinfo  {journal} {Scientific
  reports}\ }\textbf {\bibinfo {volume} {5}},\ \bibinfo {pages} {1} (\bibinfo
  {year} {2015})}\BibitemShut {NoStop}%
\bibitem [{\citenamefont {Chen}\ and\ \citenamefont
  {Goan}(2016)}]{chen2016effects}%
  \BibitemOpen
  \bibfield  {author} {\bibinfo {author} {\bibfnamefont {C.-C.}\ \bibnamefont
  {Chen}}\ and\ \bibinfo {author} {\bibfnamefont {H.-S.}\ \bibnamefont
  {Goan}},\ }\bibfield  {title} {\bibinfo {title} {Effects of initial
  system-environment correlations on open-quantum-system dynamics and state
  preparation},\ }\href@noop {} {\bibfield  {journal} {\bibinfo  {journal}
  {Physical Review A}\ }\textbf {\bibinfo {volume} {93}},\ \bibinfo {pages}
  {032113} (\bibinfo {year} {2016})}\BibitemShut {NoStop}%
\bibitem [{\citenamefont {De~Vega}\ and\ \citenamefont
  {Alonso}(2017)}]{de2017dynamics}%
  \BibitemOpen
  \bibfield  {author} {\bibinfo {author} {\bibfnamefont {I.}~\bibnamefont
  {De~Vega}}\ and\ \bibinfo {author} {\bibfnamefont {D.}~\bibnamefont
  {Alonso}},\ }\bibfield  {title} {\bibinfo {title} {Dynamics of non-markovian
  open quantum systems},\ }\href@noop {} {\bibfield  {journal} {\bibinfo
  {journal} {Reviews of Modern Physics}\ }\textbf {\bibinfo {volume} {89}},\
  \bibinfo {pages} {015001} (\bibinfo {year} {2017})}\BibitemShut {NoStop}%
\bibitem [{\citenamefont {Kitajima}\ \emph {et~al.}(2017)\citenamefont
  {Kitajima}, \citenamefont {Ban},\ and\ \citenamefont
  {Shibata}}]{kitajima2017expansion}%
  \BibitemOpen
  \bibfield  {author} {\bibinfo {author} {\bibfnamefont {S.}~\bibnamefont
  {Kitajima}}, \bibinfo {author} {\bibfnamefont {M.}~\bibnamefont {Ban}},\ and\
  \bibinfo {author} {\bibfnamefont {F.}~\bibnamefont {Shibata}},\ }\bibfield
  {title} {\bibinfo {title} {Expansion formulas for quantum master equations
  including initial correlation},\ }\href@noop {} {\bibfield  {journal}
  {\bibinfo  {journal} {Journal of Physics A: Mathematical and Theoretical}\
  }\textbf {\bibinfo {volume} {50}},\ \bibinfo {pages} {125303} (\bibinfo
  {year} {2017})}\BibitemShut {NoStop}%
\bibitem [{\citenamefont {Buser}\ \emph {et~al.}(2017)\citenamefont {Buser},
  \citenamefont {Cerrillo}, \citenamefont {Schaller},\ and\ \citenamefont
  {Cao}}]{buser2017initial}%
  \BibitemOpen
  \bibfield  {author} {\bibinfo {author} {\bibfnamefont {M.}~\bibnamefont
  {Buser}}, \bibinfo {author} {\bibfnamefont {J.}~\bibnamefont {Cerrillo}},
  \bibinfo {author} {\bibfnamefont {G.}~\bibnamefont {Schaller}},\ and\
  \bibinfo {author} {\bibfnamefont {J.}~\bibnamefont {Cao}},\ }\bibfield
  {title} {\bibinfo {title} {Initial system-environment correlations via the
  transfer-tensor method},\ }\href@noop {} {\bibfield  {journal} {\bibinfo
  {journal} {Physical Review A}\ }\textbf {\bibinfo {volume} {96}},\ \bibinfo
  {pages} {062122} (\bibinfo {year} {2017})}\BibitemShut {NoStop}%
\bibitem [{\citenamefont {Majeed}\ and\ \citenamefont
  {Chaudhry}(2019)}]{majeed2019effect}%
  \BibitemOpen
  \bibfield  {author} {\bibinfo {author} {\bibfnamefont {M.}~\bibnamefont
  {Majeed}}\ and\ \bibinfo {author} {\bibfnamefont {A.~Z.}\ \bibnamefont
  {Chaudhry}},\ }\bibfield  {title} {\bibinfo {title} {Effect of initial
  system--environment correlations with spin environments},\ }\href@noop {}
  {\bibfield  {journal} {\bibinfo  {journal} {The European Physical Journal D}\
  }\textbf {\bibinfo {volume} {73}},\ \bibinfo {pages} {1} (\bibinfo {year}
  {2019})}\BibitemShut {NoStop}%
\bibitem [{\citenamefont {Mirza}\ \emph {et~al.}(2021)\citenamefont {Mirza},
  \citenamefont {Zia},\ and\ \citenamefont {Chaudhry}}]{mirza2021master}%
  \BibitemOpen
  \bibfield  {author} {\bibinfo {author} {\bibfnamefont {A.~R.}\ \bibnamefont
  {Mirza}}, \bibinfo {author} {\bibfnamefont {M.}~\bibnamefont {Zia}},\ and\
  \bibinfo {author} {\bibfnamefont {A.~Z.}\ \bibnamefont {Chaudhry}},\
  }\bibfield  {title} {\bibinfo {title} {Master equation incorporating the
  system-environment correlations present in the joint equilibrium state},\
  }\href@noop {} {\bibfield  {journal} {\bibinfo  {journal} {Physical Review
  A}\ }\textbf {\bibinfo {volume} {104}},\ \bibinfo {pages} {042205} (\bibinfo
  {year} {2021})}\BibitemShut {NoStop}%
\bibitem [{\citenamefont {Ather}\ and\ \citenamefont
  {Chaudhry}(2021)}]{ather2021improving}%
  \BibitemOpen
  \bibfield  {author} {\bibinfo {author} {\bibfnamefont {H.}~\bibnamefont
  {Ather}}\ and\ \bibinfo {author} {\bibfnamefont {A.~Z.}\ \bibnamefont
  {Chaudhry}},\ }\bibfield  {title} {\bibinfo {title} {Improving the estimation
  of environment parameters via initial probe-environment correlations},\
  }\href@noop {} {\bibfield  {journal} {\bibinfo  {journal} {Physical Review
  A}\ }\textbf {\bibinfo {volume} {104}},\ \bibinfo {pages} {012211} (\bibinfo
  {year} {2021})}\BibitemShut {NoStop}%
\bibitem [{\citenamefont {Hall}(2000)}]{hall2000quantum}%
  \BibitemOpen
  \bibfield  {author} {\bibinfo {author} {\bibfnamefont {M.~J.}\ \bibnamefont
  {Hall}},\ }\bibfield  {title} {\bibinfo {title} {Quantum properties of
  classical fisher information},\ }\href@noop {} {\bibfield  {journal}
  {\bibinfo  {journal} {Physical Review A}\ }\textbf {\bibinfo {volume} {62}},\
  \bibinfo {pages} {012107} (\bibinfo {year} {2000})}\BibitemShut {NoStop}%
\end{thebibliography}%
\end{document}